\documentclass[prl, twocolumn, showpacs, reprint, nofootinbib, superscriptaddress, floatfix]{revtex4-2}

\usepackage{amsfonts}
\usepackage{amssymb}
\usepackage{amsmath}
\usepackage{graphics}
\usepackage{graphicx}
\usepackage{soul}
\usepackage{natbib}
\usepackage{bm}
\usepackage{mathtools}
\usepackage{stmaryrd}
\usepackage{epstopdf}
\usepackage{color}
\usepackage{lipsum}
\usepackage{balance}
\usepackage{enumitem}
\usepackage{verbatim}

\DeclareMathOperator{\sinc}{sinc}
\usepackage[colorlinks=true,citecolor=black,linkcolor=black,urlcolor=black,filecolor=black]{hyperref}

\usepackage{acronym}
\newacro{DF}{distribution function}
\newacroplural{DF}[DFs]{distribution functions}
\newcommand{\DF}{\ac{DF}}
\newacro{LDP}{Large Deviation Principle}
\newcommand{\LDP}{\ac{LDP}}
\newacro{LD}{Large Deviation}
\newcommand{\LD}{\ac{LD}}
\newacro{LDT}{Large Deviation Theory}
\newcommand{\LDT}{\ac{LDT}}
\newacro{SM}{Supplemental Material}
\newcommand{\SM}{\ac{SM}}

\DeclareMathSymbol{\shortminus}{\mathbin}{AMSa}{"39}

\newcommand{\p}{\partial}
\newcommand{\half}{\tfrac{1}{2}}
\newcommand{\rd}{\mathrm{d}}
\newcommand{\rs}{\mathrm{s}}
\newcommand{\rw}{\mathrm{w}}
\newcommand{\rT}{\mathrm{T}}

\newcommand{\deltaD}{\delta_{\mathrm{D}}}
\newcommand{\ba}{\mathbf{a}}
\newcommand{\bw}{\mathbf{w}}
\newcommand{\bwp}{\mathbf{w}^{\prime}}
\newcommand{\bW}{\mathbf{W}}
\newcommand{\bT}{\boldsymbol{\theta}}
\newcommand{\bTp}{\boldsymbol{\theta}^{\prime}}
\newcommand{\bv}{\mathbf{v}}
\newcommand{\bvp}{\mathbf{v}^{\prime}}
\newcommand{\bu}{\mathbf{u}}
\newcommand{\bk}{\mathbf{k}}
\newcommand{\bV}{\mathbf{V}}
\newcommand{\bkp}{\mathbf{k}^{\prime}}
\newcommand{\bZero}{\mathbf{0}}
\newcommand{\Mtot}{M_{\mathrm{tot}}}
\newcommand{\rNB}{\mathrm{NB}}
\newcommand{\sB}{\mathsf{B}}
\newcommand{\sQ}{\mathsf{Q}}
\newcommand{\sP}{\mathsf{P}}

\newcommand{\sZero}{\mathsf{0}}
\newcommand{\sI}{\mathsf{I}}
\newcommand{\Tdyn}{t_{\mathrm{dyn}}}
\newcommand{\Trelax}{t_{\mathrm{relax}}}
\newcommand{\mH}{\mathcal{H}}
\newcommand{\sA}{\mathsf{A}}

\newcommand{\bbP}{\mathbb{P}}
\newcommand{\bbZ}{\mathbb{Z}}
\newcommand{\mO}{\mathcal{O}}
\newcommand{\mN}{\mathcal{N}}
\newcommand{\sD}{\mathsf{D}}
\newcommand{\sR}{\mathsf{R}}
\newcommand{\re}{\mathrm{e}}
\newcommand{\ri}{\mathrm{i}}
\newcommand{\bb}{\mathbf{b}}
\newcommand{\tp}{t^{\prime}}

\newcommand{\bbR}{\mathbb{R}}
\newcommand{\Ekin}{E_{\mathrm{kin}}}
\newcommand{\Htot}{H_{\mathrm{tot}}}
\newcommand{\Etot}{E_{\mathrm{tot}}}
\newcommand{\bPtot}{\mathbf{P}_{\mathrm{tot}}}
\newcommand{\kmin}{k_{\mathrm{min}}}
\newcommand{\kmax}{k_{\mathrm{max}}}

\newcommand{\tmin}{t_{\mathrm{min}}}
\newcommand{\tmax}{t_{\mathrm{max}}}
\newcommand{\DT}{\texttt{DT}}
\newcommand{\NV}{\texttt{NV}}
\newcommand{\DV}{\texttt{DV}}
\newcommand{\VMAX}{\texttt{VMAX}}
\newcommand{\EPS}{\texttt{EPS}}
\newcommand{\omegaR}{\omega_{\mathrm{R}}}
\newcommand{\DW}{\Delta\! \bW}
\newcommand{\bDelta}{\boldsymbol{\Delta}}

\newcommand{\bSigma}{\mathbf{\Sigma}}
\newcommand{\kperp}{k_{\perp}}

\newcommand{\bA}{\mathbf{A}}

\newcommand{\Tfin}{t_\mathrm{fin}}
\newcommand{\mHI}{\mH_{\mathrm{I}}}
\newcommand{\mHEP}{\mH_{\mathrm{EP}}}
\newcommand{\mHNaive}{\mH_{\mathrm{Naive}}}
\newcommand{\eps}{\epsilon}

\newcommand{\veps}{\varepsilon}
\newcommand{\bVexact}{\mathbf{V}_{\mathrm{exact}}}
\newcommand{\tF}{\widetilde{F}}
\newcommand{\bVnum}{\mathbf{V}_{\mathrm{num}}}

\newcommand{\Tr}{\mathrm{Tr}}
\newcommand{\hbu}{\widehat{\mathbf{u}}}
\newcommand{\RePart}{\mathrm{Re}}
\newcommand{\ImPart}{\mathrm{Im}}
\newcommand{\bS}{\mathbf{S}}
\newcommand{\LangevinNaive}{\textsc{Langevin-Naive}}
\newcommand{\LangevinEP}{\textsc{Langevin-EP}}

\usepackage{soul} \usepackage{xcolor}

\begin{document}

\title{From Landau Equation and Large Deviations
\\
to Efficient Simulations of Dynamical Fluctuations}

\author{Anwar El Rhirhayi}
\affiliation{Institut Denis Poisson, UMR 7013, Universit\'{e} d’Orl\'{e}ans, Universit\'{e} de Tours, CNRS; Orl\'{e}ans, France}
\affiliation{Institut d’Astrophysique de Paris, UMR 7095, 98 bis Boulevard Arago, F-75014 Paris, France}

\author{Jean-Baptiste Fouvry}
\affiliation{Institut d’Astrophysique de Paris, UMR 7095, 98 bis Boulevard Arago, F-75014 Paris, France}

\author{Julien Barr\'{e}}
\affiliation{Institut Denis Poisson, UMR 7013, Universit\'{e} d’Orl\'{e}ans, Universit\'{e} de Tours, CNRS; Orl\'{e}ans, France}

\begin{abstract}
The (deterministic) Landau equation captures the mean long-term evolution
of dynamically hot long-range interacting finite-$N$ systems.
Though successful, this kinetic equation fundamentally ignores dynamical fluctuations.
Building upon Large Deviation Theory,
we present a physically-consistent system of Langevin equations
that simultaneously recovers the mean Landau dynamics and accurately captures the corresponding fluctuations among different realizations.
We show in particular how these Langevin equations
can be derived from Rostoker's principle
in the limit of weak two-body deflections.
We extensively validate these equations against tailored direct
$N$-body simulations, showing an exquisite level of agreement.
\end{abstract}
\maketitle

\medskip
\noindent{\em Introduction} ---
Describing the long-term dynamics of long-range interacting 
$N$-body systems is a fundamental challenge in statistical physics~\cite[see, e.g.\@,][]{Ruffo+2014}.
Such long-range dynamics cover quite a wide class of systems like plasmas~\cite{Nicholson1992}, galaxies~\cite{BinneyTremaine2008}, globular clusters~\cite{Heggie2003}, two dimensional geophysical vortices~\cite{Bouchet+2012} and many others~\citep[see, e.g.\@,][]{Dauxois+2002}. The instantaneous state of one such realization
can be tracked via its empirical distribution function: it encodes the complete information about the particles' positions in phase space~\cite[see, e.g.\@,][]{Klimontovich1967,Balescu1997}.
Classical kinetic theories focus on the dynamics
averaged over the initial conditions~\cite[see, e.g.\@,][]{Landau+1981, Fouvry+2024}.
In particular, for statistically homogeneous
and dynamically hot systems,
i.e.\ with weak collective amplification,
the ensemble-averaged empirical distribution
follows the Landau equation~\cite{Landau1936, Landau+1981}.
It describes the system's slow relaxation.

The Landau equation is, by design, deterministic.
As such, this kinetic equation is unable
to predict the behavior of typical or rare dynamical fluctuations
away from the mean evolution.
Understanding such fluctuations is both a fundamental question
as well as of prime practical importance,
e.g.\@, to capture the intrinsic variability in the time
of core collapse of globular clusters
in galactic dynamics~\citep{Lynden+1968}.
This variability is also evident in galactic discs,
where strong differences between different realizations motivate a large deviation approach~\citep[see, e.g.\@,][]{Roule+2025,Asano+2026}

Extending kinetic theories to also capture fluctuations
is the realm of \LDT\@~\citep[see, e.g.\@,][]{Touchette2009}.
In the limit ${ N \!\gg\! 1 }$, \LDT\@ provides
a quantitative description of the full statistics
of fluctuations around the mean evolution.
There has been striking progress recently
in the derivation of a \LDP\@ for the long-term
relaxation of long-range interacting systems:
it was obtained for the homogeneous Landau equation
in~\citep{Feliachi+2021},
with subsequent developments
to account for collective effects~\citep{Feliachi+2022}
or inhomogeneity~\citep{Feliachi+2024}.

In practice, exploiting these \LDP\@ is difficult, since they require the computation of a supremum in a functional space (see Eq.~\ref{eq:LDP} below).
On the other hand, characterizing fluctuations from direct $N$-body simulations
is computationally prohibitive,
because of the large scale separation between the dynamical time and relaxation time.
It is therefore highly desirable to devise numerical methods able to generate efficiently individual realizations of the dynamics, whose statistical properties are compatible with the theoretical LDP.
In~\cite{Fontbona+2009,Fu+2025}, a set of energy-preserving stochastic equations was proposed for the Coulomb case to maintain macroscopic invariants during simulation. However, the fundamental question remains: do these effective Langevin processes truly represent the macroscopic $N$-body fluctuations? In the present study, starting from the physics of two-body encounters underpinning the Landau equation, we derive a set of coupled Langevin equations.
We show, both analytically and numerically,
that these Langevin equations are the stochastic counterpart to the \LDP\@ governing the long-term relaxation. We implement them explicitly in a simple 2D model, and we show that the fluctuations they generate are in excellent agreement with those of direct $N$-body simulations. Overall, this provides --- for the first time --- a direct numerical validation of the power of \LDT\@ to study the dynamical fluctuations of long-range interacting systems.

\medskip
\noindent{\em System} ---
We consider the long-term relaxation
of a long-range interacting system in dimension $d$, comprising $N$ particles
of individual mass ${ m \!=\! \Mtot / N }$,
with $\Mtot$ the system's total mass.
We denote the phase space coordinates by
${ \bw \!=\! (\bT , \bv) }$,
assuming that the angle $\bT$ is ${2\pi}$-periodic.
In addition to their kinetic energy,
particles interact with one another through
a (truncated) gravitational potential,
${ U(\bT , \bTp) \!=\! \sum_{\bk} \psi_{\bk} \re^{\ri \bk \cdot (\bT - \bTp)} }$,
with ${ \psi_{\bk} = - G / |\bk|^{2} }$
for ${ \kmin \!\leq\! |\bk| \!\leq\! \kmax }$.
We refer to the \SM\@~\cite{SupplMat}
for more details on the precise numerical setup.

In practice, we assume that the system is
(i) dynamically hot, in order to ignore collective effects~\cite[see, e.g.\@,][]{Chavanis2013};
(ii) of infinite extent, i.e.\ ${ \kmin \!\gg\! 1 }$,
so as to simplify the collision kernel (see \SM\@~\cite{SupplMat});
(iii) statistically homogeneous, so that on long timescales
we may average over the position variables.
The instantaneous state of the system in velocity space
is described by the empirical \DF\@
\begin{equation}
F_N (\bv , t) := \sum_{i} m \, \deltaD [\bv \!-\! \bv_{i} (t)] ,
\label{def_F_N}
\end{equation}
with $\deltaD$ the Dirac delta and ${ \bv_{i} (t) }$ the velocity of particle $i$ at time $t$.
Importantly, $F_{N}$ is a stochastic quantity:
it differs from one realization to another.

\medskip
\noindent{\em Landau equation and Large deviations} ---
We assume some timescale separation between (i) the (fast) dynamical time $\Tdyn$,
i.e.\ the typical crossing time in the system,
(ii) the (slow) relaxation time ${ \Trelax \!:=\! N \, \Tdyn }$,
on which the irreversible relaxation of the system occurs. The slow dynamics is driven by two-body encounters, and
the long-term dynamics of the ensemble-averaged \DF\@, ${ F \!:=\! \langle F_{N}\rangle }$,
is described by the homogeneous Landau equation~\citep{Landau1936,Landau+1981,Chavanis2013}
\begin{equation}
\frac{\p F(\bv, t)}{\p t} \!=\! \frac{m}{2} \frac{\p}{\p \bv} \cdot \!\!\int\!\! \rd \bvp \sB(\bv \!-\! \bvp) \!\bigg[\! \frac{\p F}{\p \bv} F(\bvp) \!-\! F(\bv) \frac{\p F}{\p\bvp} \!\bigg] ,
\label{eq:Landau}
\end{equation}
where we dropped the time dependence of the \DF\@ for clarity. Here, ${ \sB }$ is the collision kernel (see \SM\@~\cite{SupplMat} for its full expression).
It encodes the geometric properties of the pairwise (weak) scatters,
in particular the conservation of energy and momentum. 
In the present context, the collision kernel takes the form
\begin{equation}
\sB (\bu) = f (u) \, \sP(\bu) ,
\label{eq:collision_kernel}
\end{equation}
where ${\sP (\bu) \!:=\! u^2 \sI \!-\! \bu \otimes \bu }$ is proportional to the orthogonal projector onto the space perpendicular to $\bu$. The scalar prefactor is given by ${ f (u) \!:=\! \kappa / u^3 }$, with $\kappa$ being a constant proportional to $G^2$.
Equation~\eqref{eq:Landau} only describes the dynamics of the mean $F$, obtained through an ensemble-average over realizations: its dynamics is slow, as evidenced from the prefactor ${m \!=\! \Mtot/N}$ in Eq.~\eqref{eq:Landau}.
It does not provide any information
on the statistics of the fluctuations of a given realization $F_{N}$
around $F$.
Leveraging the \LDT\@ for slow-fast systems~\cite{BouchetVanden2016}, \cite{Feliachi+2021} obtained an \LDP\@ to describe these fluctuations.
More precisely,
the probability that ${ F_N(\tau) }$ follows a prescribed realization ${ \tF(\tau) }$
over the time interval ${ \tau \!\in\! [0,T] }$ (with ${ T \!\sim\! \Trelax }$) can be estimated as\footnote{We also assume that ${ F_N(\tau \!=\! 0) \!\to\! \tF(\tau \!=\! 0) }$ as ${ N \!\to\! +\infty }$.}
\begin{align}
& {} \bbP (\{F_N(\tau)\}_{0\leq \tau \leq T} \!\approx\! \{\tF(\tau)\}_{0\leq \tau \leq T}) \underset{\frac{\Mtot}{m} \to +\infty}{\asymp}
\label{eq:LDP}
\\
& {} \exp \!\bigg[\! -\frac{\Mtot}{m} \sup_P \!\! \int_0^T \!\!\! \rd \tau \bigg\{ \bigg( \!\int\!\! \rd \bv \frac{\dot{\tF}(\bv)}{\Mtot} P(\bv) \bigg) \!-\! \mH[\tF, P] \bigg\} \bigg] .
\nonumber
\end{align}
In this expression, the notation ${ \asymp_{N \to + \infty} }$ denotes logarithmic equivalence in the large-$N$ limit,
while ${ P(\bv, \tau) }$ is a field conjugate to $\tF$,
acting as an effective force to enforce
the prescribed path ${ \{\tF(\tau)\}_{0\leq \tau \leq T} }$~\cite{Roma2015}. Finally, ${\mH [\tF,P] }$ denotes the \LD\@ Hamiltonian: it encodes all the statistics of the fluctuations around the mean \DF\@~\cite{Roma2015,Touchette2009}. 
Its full expression in the Landau case is provided in the \SM\@~\cite{SupplMat}.
Formally, Eq.~\eqref{eq:LDP}
provides a complete theoretical picture
of the dynamical large deviations around the Landau equation.
Yet, it is hard to use for practical computations.
Nonetheless, it suggests the possibility to design innovative numerical methods which (i) have statistical properties compatible with Eq.~\eqref{eq:LDP}; (ii) operate directly on the slow time scale $\Trelax$, generating huge efficiency gains. This is the goal of the Langevin approach detailed below.

\medskip
\noindent{\em Naive Langevin process} ---
The Landau equation~\eqref{eq:Landau} can be seen as a non linear Fokker-Planck equation. As such, one can approximate $F$
by $N$ particles evolving through coupled Langevin equations~\cite{Risken1989}.
The associated equations read, using Stratonovich formulation,
\begin{equation}
\rd \bv_{i} = \sqrt{\sum_{j \neq i} m^2 \sB(\bv_i \!-\! \bv_j)} \circ \rd \bW_{i, t} ,
\label{eq:Langevin_naive}
\end{equation}
where ${ \{ \rd \bW_{i, t}\}_{1\leq i \leq N} }$ are $N$-independent $d$-dimensional Wiener processes following ${ \rd \bW_{i, t} \!\sim\! \mN(0, \rd t \, \sI) }$,
with $\sI$ the ${ d \!\times\! d }$ identity matrix.
We refer to the \SM\@~\cite{SupplMat} for the derivation of Eq.~\eqref{eq:Langevin_naive}
and for its It\^{o} version.
In the following, we refer to this formulation as \LangevinNaive\@,
since it reproduces the correct diffusion structure of Eq.~\eqref{eq:Landau} but does not capture the underlying physics of binary deflections and the associated conservation constraints.

Simulating such Langevin equations typically offers a major computational advantage over deterministic approaches.
Indeed, it allows in principle for a much larger integration time step $\Delta t$, of the order of a small fraction of $\Trelax$~\citep[see, e.g.\@,][]{Heyvaerts+2017}.

We now test whether \LangevinNaive\@ reproduces the statistics of the $N$-body trajectories. Our initial conditions for our benchmark case are:
(i) homogeneous in space;
(ii) anisotropic Gaussian in velocities, with ${ T_x \!=\! 3 \, T_y }$. We track the system's slow relaxation toward an isotropic Gaussian state by monitoring the evolution of the kinetic temperatures $T_x$ and $T_y$, defined as
\begin{equation}
T_x(t) = \!\!\int\!\! \rd \bv \, F_N(\bv, t) (v^x)^2 = \sum_{i} m (v^x_i)^2,
\label{eq:def_Tx}
\end{equation}
with an analogous definition for ${ T_y(t) }$.
Further details concerning the $N$-body simulations and the numerical integration of \LangevinNaive\@ are available in the \SM\@~\cite{SupplMat}.

\begin{figure}[htbp!]
\centering
\includegraphics[width=0.45 \textwidth]{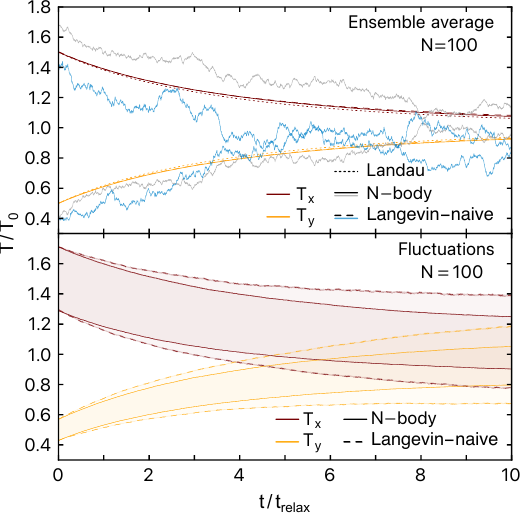}
\caption{Ensemble average (top) and 16\%--84\% fluctuations (bottom)
of the temperatures during the relaxation of ${ 10\,000 }$ systems comprising ${ N \!=\! 100 }$ particles,
as given by $N$-body simulations
and \LangevinNaive\@ (Eq.~\ref{eq:Langevin_naive}). Here $T_0$ is the average temperature at equilibrium.
In the top panel, we also show the prediction of the Landau equation
along with one realization for $N$-body and \LangevinNaive\@.
The \LangevinNaive\@ process correctly reproduces the average behavior,
but yields wrong deviations.
}
\label{fig:Langevin_naive_contours}
\end{figure}

The ensemble-averaged temperatures $T_x$ and $T_y$ for \LangevinNaive\@ and $N$-body are shown in the top of Fig.~\ref{fig:Langevin_naive_contours} and coincide with the Landau solution (see \SM\@~\cite{SupplMat} for details on its numerical integration).
The bottom panel of Fig.~\ref{fig:Langevin_naive_contours} illustrates the 16\%--84\% contours
(i.e.\ the 16th and 84th percentiles over the realizations).
It shows a clear disagreement between the $N$-body and \LangevinNaive\@ simulations.
Hence, although \LangevinNaive\@ (Eq.~\ref{eq:Langevin_naive})
correctly reproduces the average behaviors,
it fails to capture the typical deviations.
This is not surprising since \LangevinNaive\@
does not faithfully encode the underlying deflection physics.
For example, it conserves the total momentum and kinetic energy only on average, and not for each realization.

\medskip
\noindent{\em Rostoker principle and Langevin equation} ---
There are infinitely many Langevin processes corresponding to the deterministic Landau equation~\eqref{eq:Landau}: we shall now make use of this degree of freedom. Our strategy to derive a consistent effective Langevin equation is to start from the physical principles underlying the Landau equation.
When two particles ${ (i,j) }$, with velocities ${ (\bv_i,\bv_j) }$, interact, they incur deflections ${ (\delta \bv_i^j,\delta \bv_j^i \!=\! -\delta \bv_i^j) }$. Neglecting the effect of the ${ (N \!-\! 2) }$ other particles, these deflections can be analyzed directly using Hamiltonian dynamics.
They are small (unless a rare close encounter happens), and they are random variables, with a law depending on ${ (\bv_i,\bv_j) }$.
Here, the randomness comes from the (random) pre-collision spatial positions of the particles, assumed to be uniform and independent. 
Now, Rostoker's principle~\citep{Rostoker1964a,Rostoker1964b} states that the system's evolution can be described as the uncorrelated superposition of such two-body deflections. During a time interval
${ \Tdyn \!\ll\! \Delta t \!\ll\! \Trelax }$, velocities are approximately constant, and each pair ${ (i,j) }$ effectively interacts many times. By the Central Limit Theorem, ${ \Delta \bv_i^j }$, the total deflection induced by $j$ on $i$, is then a Gaussian random variable, whose mean and variance can be computed and depend on ${ (\bv_i,\bv_j) }$. Summing the contribution of all pairs, one obtains in the continuous time limit ${ \Delta t \!\to\! \rd t}$,
the following $N$ weakly coupled Langevin equations,
in Stratonovich formulation,
\begin{equation}
\rd \bv_{i} = \sum_{j \neq i} \bigg[ m \, \sB^{1/2} (\bv_i\!-\!\bv_j) \circ \rd \bW_{ij, t} \bigg] ,
\label{eq:EP_Langevin}
\end{equation}
where $\sB^{1/2}$ is a square root of the collision kernel $\sB$ and ${ \rd \bW_{ij, t} \!\sim\! \mN(0, \rd t \, \sI) }$ are ${ N(N \!-\! 1) }$ $d$-dimensional Wiener processes, which are correlated by the antisymmetry relation
\begin{equation}
\rd \bW_{ij, t} = - \rd \bW_{ji, t} .
\label{eq:Noise_antisymmetry}
\end{equation}
Without this antisymmetry constraint, Eq.~\eqref{eq:EP_Langevin} is actually strictly equivalent to \LangevinNaive\@ from Eq.~\eqref{eq:Langevin_naive}.
We refer to \SM\@~\cite{SupplMat} for details on this derivation and for the It\^{o} version of Eq.~\eqref{eq:EP_Langevin}. 
It is remarkable that this derivation from Rostoker's principle yields the same Langevin
process recently analyzed in details by~\cite{Fu+2025} for the Coulomb case, and which was, to our knowledge, originally introduced in~\cite{Fontbona+2009}.
Equation~\eqref{eq:EP_Langevin} exactly preserves the total momentum and kinetic energy.
Hence, we call it \LangevinEP\@,
with ``\textsc{EP}'' standing for ``Energy-Preserving''.
Indeed the diffusion term satisfies
\begin{equation}
\sB^{1/2} (\bv_i \!-\! \bv_j) ( \bv_i \!-\! \bv_j ) = \bZero,
\end{equation}
which means that the ``intrinsic'' stochastic term is added only along some privileged directions in velocity space. Geometrically, these directions are precisely those that are compatible with the conservation of the kinetic energy. See \SM\@~\cite{SupplMat} for more details, following on the steps from~\cite{Fu+2025}.

Crucially, beyond preserving macroscopic invariants, we analytically prove that the \LDP\@ for the empirical velocity distribution of Eq.~\eqref{eq:EP_Langevin} is governed by the exact same \LD\@ Hamiltonian as the deterministic Landau equation, thus ensuring the correct macroscopic fluctuations. To demonstrate this, we compute the \LD\@ Hamiltonian $\mathcal{H}[F, P]$ directly from its definition as the scaled cumulant generating function over an infinitesimal macroscopic time step ${\rd \tau := (m/\Mtot) \rd t}$ \cite{Feliachi+2021},
\begin{align}
\exp \!\bigg(\! \frac{\Mtot}{m} \mH[F, P] \rd \tau \!\bigg)  \!\!\! \underset{\frac{\Mtot}{m} \to \infty}{\asymp} \!\!\!
\bigg\langle\! \! \exp \!\bigg\{ \! \frac{1}{m} \!\!\int\! & {} \rd \bv \, P(\bv) \rd_\tau F_N(\bv) \!\bigg\} \!\bigg\rangle ,
\label{eq:LDP_H}
\end{align}
where ${ P(\bv) }$ is a conjugate test function probing density fluctuations, ${ \rd_\tau F_N }$ denotes the It\^{o} stochastic differential over the macroscopic time step ${ \rd \tau }$ and ${ \langle \cdot \rangle }$ denotes the expectation over the intrinsic noise.
Using the empirical measure, the argument of the exponential simplifies to a discrete sum, ${ \frac{1}{m} \!\int\! \rd \bv  P(\bv) \rd_{\tau} F_N(\bv) \!=\! \sum_i \rd P(\bv_i) }$.
Applying It\^{o}'s lemma to expand ${ \rd P(\bv_i) }$ yields a deterministic drift, a stochastic noise term, and a quadratic It\^{o} correction. The Gaussian expectation of the noise is exactly evaluated by exploiting the antisymmetry ${ \rd\bW_{ij} \!=\! -\rd\bW_{ji} }$ to group the stochastic increments into strictly independent pairs.
Overall, in the continuous large-$N$ limit, the sum of these contributions yields
\begin{equation}
\mHEP[F, P] = \mH^{(1)}[F, P] + \mH^{(2)}[F, P] ,
\label{eq:mH_EP}
\end{equation}
where the first contribution is linear in $P$,
\begin{align}
\mH^{(1)}[F, P] = {} & \half \!\!\int \!\! \rd \bv \rd \bvp \, P(\bv)
\\
& {} \times \frac{\p}{\p \bv} \!\cdot\! \bigg\{\sB(\bv \!-\! \bvp) \bigg[ \frac{\p F}{\p \bv} F(\bvp) \!-\! \frac{\p F}{\p \bvp} F(\bv) \bigg] \bigg\} ,
\nonumber
\end{align}
and the second contribution is quadratic in $P$,
\begin{align}
\mH^{(2)}[F, P] = {} & \half \!\!\int \!\! \rd \bv \rd \bvp \, \bigg[ \frac{\p P}{\p \bv} \bigg]^{\rT} \!\sB(\bv \!-\! \bvp) \bigg[ \frac{\p P}{\p \bv} \!-\! \frac{\p P}{\p \bvp} \bigg]
\nonumber
\\
\times \, & {}  F(\bv) F(\bvp) .
\end{align}
This exactly matches the \LD\@ Hamiltonian derived for the deterministic Landau equation~\cite{Feliachi+2021}.
It proves that \LangevinEP\@ inherently captures the correct fluctuation statistics. The full analytical derivation is detailed in the \SM\@~\cite{SupplMat}.
Importantly, the cross-term in ${\p_{\bv}P \p_{\bvp} \!P }$
in ${\mH^{(2)}}$ captures the pairwise correlations that \LangevinNaive\@ misses.

\medskip
\noindent{\em Validation} ---
We now turn to the numerical implementation of \LangevinEP\@, using the approach and algorithm
developed in~\cite{Fu+2025}.
The Stratonovich formulation~\eqref{eq:EP_Langevin} has no drift term, hence suggesting the use of the implicit midpoint method. However, due to the singular form of ${ \sB(\bu) }$ (see Eq.~\ref{eq:collision_kernel}), this numerical scheme fails to converge~\cite{Fu+2025}. 
Following the same approach as in~\cite{Fu+2025}, we use a modified version of the fully implicit midpoint method which (i) converges appropriately to \LangevinEP\@~\eqref{eq:EP_Langevin}; (ii) is explicitly solvable due to its linear dependence on the implicit variable and (iii) exactly conserves energy and momentum for finite time steps (up to round-off errors).
In our simulations, we also deliberately restrict our integration to a small time step (${\Delta t \sim \Tdyn}$) to ensure that the numerical scheme does not contaminate the delicate signal of the fluctuation statistics.
All the details about the algorithm and the numerical scheme are presented in the \SM\@~\cite{SupplMat}.

\begin{figure}[htbp!]
\centering
\includegraphics[width=0.45 \textwidth]{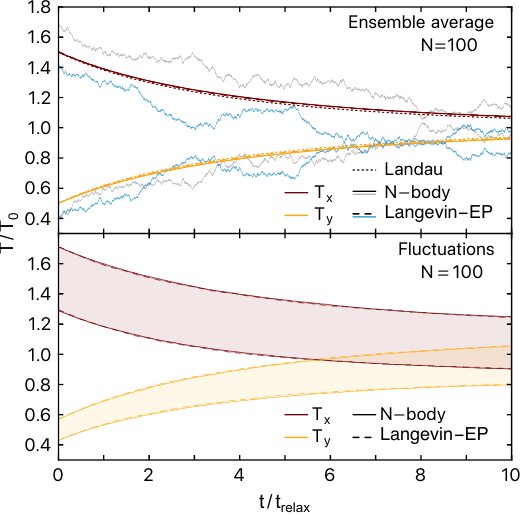}
\caption{
Ensemble average and fluctuations contours of the relaxation process over ${ 10\,000 }$ \LangevinEP\@ and $N$-body realizations. In the top panel, we show the ensemble average as well as the prediction given by the Landau equation. A single realization of both \LangevinEP\@ and $N$-body simulations is also displayed. In the bottom panel, we show the contours at 16\%--84\%.
\LangevinEP\@ not only reproduces the average behavior, but also yields the correct deviations.
}
\label{fig:Money_plot}
\end{figure}

Armed with this numerical algorithm,
we can compare the statistics of the typical fluctuations of \LangevinEP\@ with direct $N$-body numerical simulations.
We use the same test case as for \LangevinNaive\@.
The average temperature obtained with \LangevinEP\@ and the $N$-body simulations is displayed in the top panel of Fig.~\ref{fig:Money_plot}.
Both results agree with the Landau solution, up to a small offset visible between the three curves.
This discrepancy originates from finite-$N$ effects and higher-order correlations neglected by the kinetic theory. We confirm in the \SM\@~\cite{SupplMat} that this offset vanishes as $\mathcal{O}(1/N)$, ensuring that all three descriptions coincide in the large-$N$ limit.
The bottom panel of Fig.~\ref{fig:Money_plot} illustrates the 16\%--84\% contours and shows a remarkable agreement between the $N$-body and \LangevinEP\@ simulations.
This is the main result of this work.
In particular, we emphasize that the match observed in Fig.~\ref{fig:Money_plot}
was by no means guaranteed.
In the \SM\@~\cite{SupplMat}, we consider a waterbag initial condition and recover the same level of agreement.

\begin{figure}[htbp!]
\centering
\includegraphics[width=0.45 \textwidth]{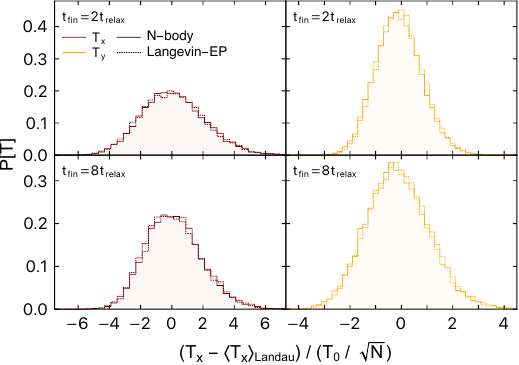}
\caption{Histogram representing the temperature distribution at ${ t \!=\! 2 \, \Trelax }$ and ${ t \!=\! 8 \Trelax }$ for ${ N \!=\! 100 }$ particles and over ${ 10\,000 }$ realizations. The \LangevinEP\@ process \eqref{eq:EP_Langevin} agrees with the distribution given by the $N$-body simulations.
}
\label{fig:histograms}
\end{figure}

For a more detailed comparison,
we present in Fig.~\protect\ref{fig:histograms} some histograms of $T_x$ and $T_y$.
Here as well, very good agreement is observed between the \LangevinEP\@ and $N$-body histograms, despite the non-Gaussian features of the probability distribution functions. All these results (i) suggest that Rostoker's principle is the right approach to mimic the microscopic $N$-body dynamics; (ii) validate numerically the statistics of typical fluctuations and (iii) reinforce the idea that \LangevinEP\@ (Eq.~\ref{eq:EP_Langevin}) captures correctly the essence of the $N$-body dynamics, by mimicking the deterministic dynamics with some ``intrinsic'' and directionally-filtered noise.

In addition to the temperature profiles, in the \SM\@~\cite{SupplMat} we present the contours of some cubic observables.
As for the temperature, \LangevinEP\@ and $N$-body simulations exhibit close agreement.

\medskip
\noindent{\em Conclusion and outlook} ---
In the present work, we have derived and implemented
an efficient effective stochastic method able to mimic the slow relaxation of a self-gravitating system, including the statistics of finite-$N$ fluctuations.
Comparison between the effective method and direct $N$-body simulations shows a near perfect agreement. This opens the door to huge gains in the numerical simulation of collisional relaxation.

This success suggests that, beyond the effective Langevin dynamics used here,
several alternative strategies could be considered to tackle the issue of fluctuations during collisional relaxation. 
First, by picking the colliding pairs randomly,~\cite{Kai+2025}
could reproduce these stochastic effects at a reduced computational cost.
Second, integrating the deterministic Lyapunov hierarchy~\cite{Nicholson1992,Bouchet+2013} offers a direct route to compute the equal-time correlation functions without relying on stochastic sampling. Third, recasting \LangevinEP\@ into a Dean--Kawasaki framework~\cite{Dean1996, Kawazaki1998} could provide a completely different approach to capturing both small and large fluctuations~\cite{Cornalba+2023}.

Finally, in this Letter we restricted ourselves to high dynamical temperatures. Outside of this regime, \LangevinEP\@ fails to accurately describe fluctuation statistics (see \SM\@~\cite{SupplMat}). In this case, the average relaxation is ruled by the (homogeneous) Balescu--Lenard equation~\cite{Balescu1960,Lenard1960}, and the \LD\@ Hamiltonian becomes non-quadratic in the conjugate field~\cite{Feliachi+2022}.
The generalization of Eq.~\eqref{eq:EP_Langevin}
to this case is completely open.

\bigskip
\begin{acknowledgments}
This work is partially supported by the grants SEGAL ANR-19-CE31-0017, GALBAR ANR-25-CE 31-4684, BEYOND-BL ANR-25-CE57-2626 and RETENU ANR-20-CE40-0005-01 of the French Agence Nationale de la Recherche.
This project has received financial support from the CNRS through the MITI interdisciplinary programs.
This work has made use of the Infinity Cluster hosted by Institut d'Astrophysique de Paris, partially funded by IDF-DIM-ORIGINES-2023-4-11. 
We thank St\'ephane Rouberol for the smooth running of the
Infinity cluster. We warmly thank C. Pichon, F. Bouchet and O. Feliachi for insightful discussions.
\end{acknowledgments}

\medskip
\noindent{\em Data availability} --- The data that support the findings of
this article are openly available~\cite{github}.

\clearpage
\setcounter{equation}{0}

\begin{center}
\textbf{End Matter}
\end{center}
\vspace{-0.3cm}
\setcounter{equation}{0}
\renewcommand{\theequation}{EM\arabic{equation}}

\section{Rostoker approach}
\label{app:Rostoker_protocol}

In this Appendix, we detail the Rostoker approach
introduced in the main text.
First, we introduce an intermediate timescale ${ \Delta  t }$ such that
\begin{equation}
\Tdyn \ll \Delta  t \ll \Trelax.
\end{equation}
Over the interval ${ [t , t \!+\! \Delta t] }$,
each pair ${ (i,j) }$ of particles undergoes ${ n \!\gg\! 1 }$ weak deflections. A key assumption of the protocol is that all these deflections are computed using the velocities at time $t$, i.e.\ the deflections occurring within ${ [t, t \!+\! \Delta t] }$ do not modify the initial data entering any other deflections.

At time $t$, the particle $i$ has velocity $\bv_i$ and at ${ t \!=\! t \!+\! \Delta  t }$, this velocity changes by an increment ${ \delta \bv_i }$, after it deflects $n$ times with each of the ${(N \!-\! 1})$ other particles simultaneously.
For a given pair, ${ (i,j) }$, the deflection labeled ${ (p) }$ lasts for ${ \Delta  t / n }$ and contributes an increment ${ \delta \bv_i^{j,(p)} }$ to particle $i$.
With this notation, the total increment,
${ \Delta \bv_{i}^{j} }$,
in the velocity of particle $i$ sourced by particle $j$
over the time interval ${ [t , t \!+\! \Delta  t] }$, is therefore
${ \Delta \bv_i^{j} \!=\! \sum_{p = 1}^n \delta \bv_i^{j,(p)} }$.
Without loss of generality, we set ${ t \!=\! 0 }$ for the remainder of this discussion.
To pursue our calculation, we introduce
a couple of additional assumptions: each deflection
(i) is independent of the previous one;
(ii) is independent from the others that happen simultaneously;
(iii) and is governed by the deterministic Hamiltonian of the two-body problem~\cite{Hamilton2021}. 
Following Eq.~{(37)} in~\cite{Heyvaerts+2017},
after a deflection ${ (i,j) }$,
the velocity increment is
\begin{equation}
\delta \bv_i^{j} = - \!\! \int_0^{\Delta  t /n} \!\!\!\!\!\! \rd \tp \, m \sum_{\bk} \ri \bk \, \re^{\ri \bk \cdot [\bT_i(\tp) - \bT_j(\tp)]} \psi_{\bk} .
\label{eq:deltav_Rostoker}
\end{equation}
where we shortened the notations with
${ \delta \bv_{i}^{j} \!=\! \delta \bv_{i}^{j,(p)} }$.
In this expression, ${ \bT_i (\tp) }$ and ${ \bT_j (\tp) }$ denote the true, interacting trajectories of the particles. To explicitly compute the drift and diffusion coefficients in the weak-coupling regime,
characteristic of the Landau equation, one must evaluate this integral via a perturbative expansion of these trajectories.
This is detailed in the \SM\@~\cite{SupplMat}.
We emphasize that Eq.~\eqref{eq:deltav_Rostoker} satisfies
the momentum conservation since
\begin{equation}
\delta \bv_i^{j} = - \delta \bv_j^{i} .
\label{eq:deltaV_momentum_cons}
\end{equation} 
At the initial time $t$, we assume that the system is well phase-mixed.
We further have ${ \Tdyn \!\ll\! \Delta  t \!\ll\! \Trelax }$
so that many weak deflections occur in the interval ${ [t, t \!+\! \Delta  t] }$
during which we assume that the empirical \DF\@ remains unchanged.
Under these conditions, the velocities ${ \bv_i (t) }$ can be approximated as a diffusion process in velocity space.
Indeed, since each deflection is independent from the previous ones,
each particle's total velocity increment ${ \Delta \bv_i^{j} }$ is the sum of ${ n \!\gg\! 1 }$ small and independent stochastic increments, ${ \delta \bv_i^{j} }$. Furthermore, the grazing collision limit ensures that these small kicks have a finite variance.
In the limit ${ n \!\gg\! 1 }$,
the Functional Central Limit Theorem~\cite{Bhattacharya2021}
implies that ${ \Delta \bv_i^{j} }$ satisfies the Langevin process
\begin{equation}
\Delta \bv_i^{j} = \bb(\bv_i, \, \bv_j) \, \Delta t + \sD^{1/2} (\bv_i, \, \bv_j) \, \DW_{ij,  t}.
\label{eq:Brownian_motion_delta_v}
\end{equation}
Here, the noise statistics is given by
\begin{subequations}
\begin{align}
\langle \DW_{ij,  t} \rangle & {} \!=\! 0 ,
\label{eq:dW_1_noise_Rostoker}
\\
\langle \DW_{ij, t} \!\otimes\! \DW_{kl, t} \rangle & {} \!=\! \sI ( \delta_{ik} \delta_{jl} \!-\! \delta_{il} \delta_{jk} ) \Delta t ,
\label{eq:dW_2_noise_Rostoker}
\end{align}
\label{eq:stat_noise_Rostoker}\end{subequations}
from which it follows that
\begin{equation}
\DW_{ij, t} = - \DW_{ji, t} .
\label{eq:antisymmetry_W_ij}
\end{equation}

The associated friction and diffusion terms are given by~\cite[see, e.g.\@,][]{Risken1989}
\begin{subequations}
\begin{align}
\bb(\bv_i, \, \bv_j) & {} = \lim_{\Delta  t \to + \infty} \langle \Delta \bv_i^{j}\rangle_{\bT} / \Delta  t ,
\label{eq:def_b_Rostoker}
\\
\sD(\bv_i, \bv_j) & {} = \lim_{\Delta t \to + \infty} \langle \Delta \bv_i^{j} \!\otimes\! \Delta \bv_i^{j}\rangle_{\bT} / \Delta t ,
\label{eq:def_D_Rostoker}
\end{align}
\label{eq:def_b_D_Rostoker}\end{subequations}
where ${ \langle \cdot \rangle_{\bT} }$ is the average over the initial phase
of the particles.
For notational simplicity, here we omit the explicit dependence on $t$.
By explicitly evaluating the microscopic velocity increments ${ \Delta \bv_i^j }$ via Eq.~\eqref{eq:deltav_Rostoker} and carrying out the phase averaging, one obtains
\begin{subequations}
\begin{align}
\bb(\bv_i, \bv_j) & {} = m^2 \p_{\bv_i} \!\cdot\! \sB(\bu_{ij}) ,
\\
\sD(\bv_i, \bv_j) & {} = m^2 \, \sB(\bu_{ij}) ,
\end{align}
\end{subequations}
where ${ \bu_{ij} \!=\! \bv_i \!-\! \bv_j }$.
The full derivation of these drift and diffusion terms is detailed in the \SM\@~\cite{SupplMat}.
Collisions from different particles $j$ are simultaneous and independent.
As a result, the total contribution to the velocity of particle $i$ is given by
${ \rd \bv_{i} \!=\! \sum_{j \neq i} \rd \bv_{i}^{j} }$, reading
\begin{equation}
\rd \bv_i \!=\! \sum_{j \neq i} \bigg[ m^2 \p_{\bv_i} \!\!\cdot\! \sB(\bu_{ij}) \rd t + m \sB^{1/2} (\bu_{ij}) \rd\bW_{ij,t} \bigg] .
\label{eq:Langevin_EP_function_of_B}
\end{equation}
It corresponds to the It\^{o} writing of Eq.~\eqref{eq:EP_Langevin} in the main text. 
We emphasize that this expression retains the exact discrete collision kernel, as we evaluate the sum over wave-vectors $\sum_{\mathbf{k}}$ directly rather than approximating it with a continuous integral (see \SM\@~\cite{SupplMat} for details).

\clearpage 

\onecolumngrid
\begin{center}
\textbf{\large Supplemental Material \\ 
From Landau Equation and Large Deviations \\
to Efficient Simulations of Dynamical Fluctuations}\\[.2cm]
Anwar El Rhirhayi$^{1,2}$, Jean-Baptiste Fouvry$^2$, and Julien Barré$^1$\\[.2cm]
{\itshape ${}^1$Institut Denis Poisson, Université d’Orléans, CNRS Université de Tours, Orléans, France\\
${}^2$Institut d’Astrophysique de Paris, UMR 7095, 98 bis Boulevard Arago, F-75014 Paris, France}
\\[1cm]
\end{center}
\twocolumngrid

\setcounter{secnumdepth}{3} 
\setcounter{section}{0}
\setcounter{subsection}{0}
\setcounter{equation}{0}
\setcounter{figure}{0}
\setcounter{table}{0}
\setcounter{page}{1}

\renewcommand{\thesection}{\Alph{section}}
\renewcommand{\thesubsection}{\arabic{subsection}}
\renewcommand{\theequation}{SM\arabic{equation}}
\renewcommand{\thefigure}{SM\arabic{figure}}

\section{\texorpdfstring{$N$}{N}-body simulations}
\label{app:NBody}

\subsection{\texorpdfstring{$N$}{N}-body system}
\label{app:NBody_system}

In this Appendix, we detail our setup for the $N$-body simulations.
All the elements detailed below are available
in an efficient \texttt{julia} code,
which is publicly distributed~\citep{github}.

As introduced in the main text, we place ourselves in two dimensions (${d \!=\! 2}$) and denote the canonical phase space coordinates with
${ \bw \!=\! (\bT, \bv) }$, where the angle $\bT$ is $2\pi$-periodic.
We stress that the position space is multi-periodic.
As detailed in Appendix~\ref{app:Collision_kernel},
by tuning the pairwise
interaction $N$-body simulations, the present multi-periodic setup
can be used to reproduce the dynamical features of a system of infinite extent.

In addition to their kinetic energy,
particles are coupled to one another
via the pairwise interaction potential
\begin{equation}
U (\bw , \bwp) = U (|\bT \!-\! \bTp|) .
\label{eq:def_U_interaction}
\end{equation}
As such, the system's total Hamiltonian is given by
\begin{equation}
\Htot = \sum_{i} \half m \bv_{i}^{2} + \sum_{\substack{i,j \\ i < j}} m^{2} U (\bT_{i} , \bT_{j}) .
\label{eq:Htot_NBODY}
\end{equation}
This Hamiltonian has two invariants,
namely
\begin{subequations}
\begin{align}
\Etot {} & = \Htot
\quad\quad\;\; \text{(Total energy)} ,
\label{eq:inv_Etot_NBODY}
\\
\bPtot {} & = \sum_{i} m \bv_{i}
\quad \text{(Total momentum)} .
\end{align}
\label{eq:inv_NBODY}\end{subequations}

Given that the angles are ${2\pi}$-periodic,
we can develop Eq.~\eqref{eq:def_U_interaction} into
\begin{equation}
U (\bT , \bTp) = \sum_{\bk \in \bbZ^{d} \backslash \{\bZero\}} \psi_{\bk} \, \re^{\ri \bk \cdot (\bT - \bTp)} .
\label{eq:Fourier_U}
\end{equation}
We take the interaction potential, ${ \psi_{\bk} \!=\! \psi_{|\bk|} }$,
to be
\begin{equation}
\psi_{\bk} = \begin{cases}
- G / |\bk|^{2}
\quad \text{for} \quad
\kmin \!\leq\! |\bk| \!\leq\! \kmax ,
\\
0
\quad \text{otherwise}.
\end{cases}
\label{eq:val_Uk}
\end{equation}
with ${ (\kmin , \kmax) }$ some (finite) given domain.
In practice, we used ${ \kmin \!=\! 10 }$
to ensure that the present multi-periodic interaction potential
can be taken as isotropic,
and imposed ${ \kmax \!=\! 100 }$, so as to prevent
any divergence of the interaction potential on small scales.

In the numerical implementation, it is convenient to split
the considered vectors $\bk$,
i.e.\ the domain ${ \kmin \!\leq\! |\bk| \!\leq\! \kmax }$,
into a positive and a negative region.
Namely, we fix the sign of a given $\bk$ according to the sign
of its first non-zero coefficient.
We note that this is unambiguous provided ${ \bk \!=\! \bZero }$ is not included in the sum from Eq.~\eqref{eq:Fourier_U}. 
Doing so, we can rewrite Eq.~\eqref{eq:Fourier_U} as
\begin{equation}
U (\bT , \bTp) = 2 \sum_{\bk > 0} \psi_{\bk} \, \cos \big[ \bk \!\cdot\! (\bT \!-\! \bTp) \big] .
\label{eq:rewrite_U_with_cos}
\end{equation}

\subsection{Equations of motion}
\label{app:NBody_EOM}

Hamilton's equations for particle $i$ read
\begin{subequations}
\begin{align}
\dot{\bT}_{i} {} & \!=\! \tfrac{1}{m} \p_{\bv_{i}} \Htot \!=\! \bv_{i} ,
\label{eq:EOM_theta_NBODY}
\\
\dot{\bv}_{i} {} & \!=\! - \tfrac{1}{m} \p_{\bT_{i}} \Htot \!=\! \sum_{j} \! \sum_{\bk > 0} \! 2 m \psi_{\bk} \bk \sin \!\big[ \bk \!\cdot\! (\bT_{i} \!-\! \bT_{j}) \!\big] ,
\label{eq:EOM_v_NBODY}
\end{align}
\label{eq:EOM_NBODY}\end{subequations}
where, importantly, we could safely include the self-interaction
contribution from ${ i \!=\! j }$. This is convenient for the numerical implementation. Using trigonometric relations,
we can rewrite Eq.~\eqref{eq:EOM_v_NBODY} into
\begin{equation}
\frac{\rd \bv_{i}}{\rd t} = \sum_{\bk > 0} 2 \psi_{\bk} \bk \big\{ \sin \big[ \bk \!\cdot\! \bT_{i} \big] C_{\bk} - \cos \big[ \bk \!\cdot\! \bT_{i} \big] \, S_{\bk} \big\} ,
\label{eq:EOM_v_mag_NBODY}
\end{equation}
where we introduced the ``magnetisations''
\begin{subequations}
\begin{align}
S_{\bk} {} & = \sum_{i} m \, \sin \big[ \bk \!\cdot\! \bT_{i} \big] ,
\label{eq:def_mag_Sk}
\\
C_{\bk} {} & = \sum_{i} m \, \cos \big[ \bk \!\cdot\! \bT_{i} \big] .
\end{align}
\label{eq:def_mag}\end{subequations}
At this stage, we note that
the magnetisations ${ (S_{\bk} , C_{\bk}) }$
are (i) shared among all the particles
and (ii) are computable in ${ \mO (N) }$ operations.
Up to the accounting of some (constant) self-interaction terms,
the same magnetisations can be used to rewrite the total energy
from Eq.~\eqref{eq:Htot_NBODY} into
\begin{equation}
\Etot = \sum_{i} \half m \bv_{i}^{2} + \sum_{\bk > 0} \psi_{\bk} \, \big\{ S_{\bk}^{2} + C_{\bk}^{2} \big\} .
\label{eq:rewrite_Etot_mag_NBODY}
\end{equation}

\subsection{Numerical integration}
\label{app:NBody_Numerical}

To evaluate the rates of change,
${ \{ (\dot{\bT}_{i} , \dot{\bv}_{i}) \}_{i} }$
(Eq.~\ref{eq:EOM_NBODY}),
we need to compute the trigonometric functions
${ \{ \sin [\bk \!\cdot\! \bT_{i}] , \cos [\bk \!\cdot\! \bT_{i}] \}_{\bk , i} }$.
These need to be computed twice:
(i) to evaluate the magnetisations, ${ (S_{\bk} , C_{\bk}) }$ (Eq.~\ref{eq:def_mag});
(ii) to evaluate the rates of change (Eq.~\ref{eq:EOM_NBODY}).
These two steps are the most taxing part of the numerical code.

In practice, we can accelerate the evaluation
of these trigonometric functions
using appropriate recurrence relations.
Given some angle
${ \bT \!=\! (\theta_{x} , \theta_{y}) }$,
we compute once
${ (\sin [\theta_{x}] , \cos [\theta_{x}]) }$
as well as ${ (\sin [\theta_{y}] , \cos [\theta_{y}]) }$.
Then, introducing ${ (s_{\bk} , c_{\bk}) \!=\! (\sin [\bk \!\cdot\! \bT] , \cos [\bk \!\cdot\! \bT])}$,
we generically have
\begin{subequations}\begin{align}
s_{\bk + \bkp} {} & = s_{\bk} c_{\bkp} + c_{\bk} s_{\bkp} ,
\label{eq:duplication_formula_sin}
\\
c_{\bk + \bkp} {} & = c_{\bk} c_{\bkp} - s_{\bk} s_{\bkp} .
\label{eq:duplication_formula_cos}
\end{align}
\label{eq:duplication_formula}\end{subequations}
Leveraging this recurrence relation, 
one can efficiently compute ${ \{ (s_{\bk} , c_{\bk}) \}_{\bk} }$
for all ${ \bk \!>\! 0 }$
that contribute to the pairwise interaction.

Once the rates of changes estimated,
we proceed with the time integration.
Fortunately, the Hamiltonian from Eq.~\eqref{eq:Htot_NBODY}
is separable, allowing us to resort
to standard symplectic methods~\citep[see, e.g.\@, Section~{V.3} in][]{Hairer+2006}.
Given some timestep \DT\@, this approach proceeds
by series of kicks and drifts.
Because the considered interaction is rather smooth,
we settled on using the sixth-order symplectic method from~\citep{Yoshida1990}.
It requires seven evaluations of the accelerations per timestep.

\begin{figure}[htbp!]
\centering
\includegraphics[width=0.43 \textwidth]{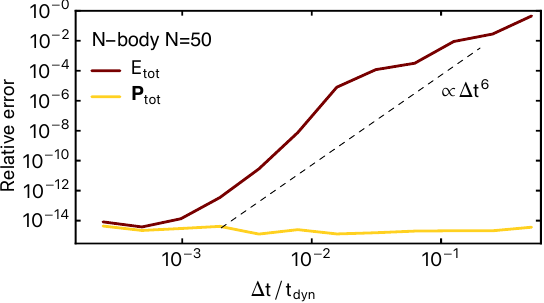}
\includegraphics[width=0.45 \textwidth]{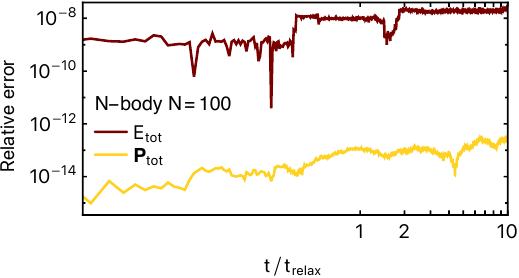}
\caption{Top: Relative error in $\Etot$ and $\bPtot$
in the $N$-body simulations (same setup as in Fig.~\ref{fig:Money_plot}),
at time ${ t \!=\! \Trelax }$,
w.r.t.\ their initial values,
as one varies the integration timestep, \DT\@.
As expected, the $N$-body integration scheme is sixth-order accurate.
Bottom: Long-time behavior of the relative errors
in $\Etot$ and $\bPtot$ in the $N$-body simulations,
w.r.t.\ their initial values.
On long timescales, the relative error in $\Etot$ is bounded
because the integration scheme is symplectic.
}
\label{fig:cv_NBODY_DT}
\end{figure}

In Fig.~\ref{fig:cv_NBODY_DT}, we verify the correct implementation
of this numerical scheme.
In that figure, we recover that (i) the finite-time error in $\Etot$ and $\bPtot$ scales like ${ \mO (\DT^6) }$;
(ii) the long-time error in $\Etot$ is bounded
owing to the symplectic nature of the integration scheme.

\subsection{Numerical parameters}
\label{app:Numerical_parameters}

In practice, to obtain the plots in Fig.~\ref{fig:Money_plot} and~\ref{fig:histograms}, we considered ${ N \!=\! 100 }$ particles.
We performed ${ 10\,000 }$ independent realizations
of the $N$-body simulation up to ${ \Tfin \!=\! 10 \, \Trelax }$,
using ${ \Delta t \!=\! 0.005 \, \Tdyn}$.

\textit{Initial conditions}. At ${ t \!=\! 0 }$,
particles are drawn independently from one another
with a uniform angle $\bT$.
Velocities are drawn from the anisotropic Maxwell distribution
\begin{equation}
F(\bv; t \!=\!0) = \Mtot \frac{1}{2 \pi} \frac{1}{\sqrt{T_{x}^{\ri} T_{y}^{\ri}}} \exp \bigg[- \frac{| v_{x} |^2}{2 T_{x}^{\ri}} -\frac{ |v_{y} |^2}{2 T_{y}^{\ri}} \bigg] ,
\end{equation}
with the initial temperatures ${ (T^{\ri}_{x} , T^{\ri}_{y}) \!=\! (1.5 \, T_{0} , 0.5 \, T_{0}) }$.
Here, ${ T_{0} \!=\! v_0^2 }$ where ${ v_0 }$ represents the typical particle velocity at equilibrium.
Two-body relaxation drives this system towards the thermodynamical equilibrium with ${ T_x \!=\! T_y \!=\! T_{0} }$.

\textit{Bootstrap}. To estimate uncertainties in Fig.~\ref{fig:Money_plot}, we use bootstraps.
More precisely, we perform the ensemble average over a sample of 10\,000 realizations,
drawn with repetitions. We perform this resampling 1\,000 times
to estimate the 16\%--84\% contours
of any observable.

\section{Collision kernel}
\label{app:Collision_kernel}

Following~\cite{Feliachi+2021, Feliachi+2024}, we introduce the collision kernel
\begin{equation}
\sB(\bv \!-\! \bvp) := 2\pi \sum_{\bk \in \bbZ^{d}} \bk \!\otimes\! \bk \, |\psi_{\bk}|^2 \, \deltaD \big[ \bk \!\cdot\! (\bv \!-\! \bvp) \big] .
\label{eq:def_B_sumK}
\end{equation}
It quantifies how a particle with velocity $\bv$ is diffusely kicked by another particle with velocity $\bvp$ through the gravitational potential $\psi_{\bk}$.
Here, $\deltaD$ is the Dirac delta, and $\bk$ the resonance vector.

In this Appendix, we compute explicitly the collision kernel $\sB$ from the expression in Eq.~\eqref{eq:def_B_sumK}. More precisely, 
we show that $\sB$ takes the form of a projector in the $\bv$ space.
Introducing the velocity difference ${ \bu \!=\! \bv \!-\! \bvp }$,
we first rewrite Eq.~\eqref{eq:def_B_sumK} as
\begin{equation}
\sB(\bu) = 2\pi \sum_{\bk \in \bbZ^d} \bk \!\otimes\! \bk \, |\psi_{\bk}|^2 \deltaD \big[ \bk \!\cdot\! \bu \big] .
\end{equation}
For practical purposes, and to enable direct comparison with the 
$N$-body simulations, we consider
\begin{equation}
\sB_{\rNB} (\bu) = 2 \pi \sum_{\mathclap{\substack{\bk \in \bbZ^{d} \\ \kmin \leq |\bk| \leq \kmax }}} \bk \!\otimes\! \bk |\psi_{\bk}|^{2} \deltaD [\bk \!\cdot\! \bu] .
\label{eq:def_sBNB}
\end{equation}
This corresponds exactly to the two-body interaction used in the simulations.
In order to fall back on the case of a system of infinite extent,
we approximate this discrete sum via
\begin{equation}
\sB_{\rNB} (\bu) \simeq 2\pi \!\! \int_{\kmin}^{\kmax} \!\!\!\! \rd \bk \, \bk \!\otimes\! \bk \, |\psi_{\bk}|^{2} \, \deltaD [\bk \!\cdot\! \bu] .
\label{eq:approx_sBNB}
\end{equation}

Such an approximation is expected to be accurate
provided that ${ \kmin \!\gg\! 1 }$,
i.e.\ provided the number of discrete resonances considered
is large.
Assuming the classical purely gravitational potential ${ \psi_{\bk} \!=\! - G / k^2 }$,
the collision kernel reads
\begin{equation}
\sB_{\rNB} (\bu) \simeq 2\pi G^2 \!\!\int_{\kmin}^{\kmax} \!\!\!\! \rd\bk \, \frac{\bk \!\otimes\! \bk}{k^4} \deltaD[\bk \!\cdot\! \bu ].
\label{eq:sigma_midpoint_impl}
\end{equation}

To evaluate this integral in an arbitrary dimension $d$, we decompose the wavevector into parallel and perpendicular components w.r.t.\ the relative velocity via ${ \bk \!=\! k_{\parallel} \hbu \!+\! \bk_{\perp} }$
with ${ \hbu \!=\! \bu/u }$. The Dirac delta function, ${ \deltaD [\bk \!\cdot\! \bu] = \deltaD [k_{\parallel}] / u }$, restricts the integration to the ${ (d \!-\! 1) }$-dimensional hyperplane orthogonal to $\bu$.

By adopting spherical coordinates for the transverse component $\bk_{\perp}$, the angular integration over the unit sphere $S^{d-2}$ becomes isotropic. Consequently, the tensorial part of the integral is proportional to the orthogonal projector ${ \sI_{\perp} = \sI - \hbu \!\otimes\! \hbu }$, with the normalization constant determined by taking the trace. The remaining radial integration on $\kperp$ yields the Coulomb logarithm $\ln \Lambda$. The collision kernel can thus be written in the compact form
\begin{equation}
\sB_{\rNB} (\bu) = f (u) \, \sP(\bu) ,
\label{eq:def_sB_generic}
\end{equation}
where 
\begin{equation}
\sP (\bu) := u^2 \sI \!-\! \bu \!\otimes\! \bu,
\label{eq:def_sP}
\end{equation} 
is proportional to the projector orthogonal to $\bu$, and the scalar prefactor is ${ f (u) \!:=\! \kappa / u^3 }$, with
\begin{equation}
\kappa := 2 \pi G^2 \ln \Lambda \, \frac{1}{d \!-\! 1}\frac{2\pi^{(d-1)/2}}{\Gamma [ \tfrac{d-1}{2} ]} .
\label{eq:def_kappa}
\end{equation}
The geometric factor $\ln \Lambda$ depends on the spatial dimension $d$ and the scales of the system via
\begin{equation}
\ln \Lambda =
\begin{cases}
\displaystyle \ln [\kmax / \kmin] {} & \text{if} \quad d = 3 ,
\\[1.0ex]
\displaystyle \big[ \kmax^{d-3} - \kmin^{d-3}\big] / (d \!-\! 3) {} & \text{otherwise} .
\end{cases}
\label{eq:lnLambda_generic}
\end{equation}

Specifically, for the two-dimensional case (${ d \!=\! 2 }$), the kernel naturally simplifies to
\begin{equation}
\sB_{\rNB}(\bu) = \frac{\kappa}{u^3}
\begin{pmatrix}
u_y^2 & -u_x u_y \\
-u_y u_x & u_x^2
\end{pmatrix} ,
\end{equation}
with ${ \kappa \!=\! 4 \pi G^2 \ln \Lambda }$ and ${ \ln \Lambda \!=\! 1/\kmin \!-\! 1/\kmax }$.

\section{Langevin-Naive}
\label{app:Naive_Langevin_simulations}

In this Appendix, we detail our effective implementation
of \LangevinNaive\@ (see Eq.~\ref{eq:Langevin_naive} in the main text).
All the elements presented in this Appendix
are implemented in an efficient \texttt{julia} code,
which is publicly distributed~\cite{github}.

\subsection{From Landau to Langevin-Naive}
\label{app:Landau_Naive_Langevin}

Following the same approach as~\cite{Feliachi+2021}, we derive \LangevinNaive\@ from the Landau equation (see Eq.~\ref{eq:Landau} in the main text) by expressing it as a self-consistent Fokker--Planck equation,
\begin{equation}
\p_t F \!=\!\frac{\p }{\p \bv} \!\cdot\! \bigg[ \!-\! \bb[F](\bv) F(\bv) + \half \frac{\p }{\p \bv} \!\cdot\! \bigg(\! \sD[F](\bv) F(\bv) \!\bigg) \bigg] ,
\end{equation}
with the drift and diffusion terms
\begin{subequations}
\begin{align}
\bb[F](\bv) & {} = m \!\!\int\!\! \rd \bvp \, F(\bvp) \p_{\bv} \!\cdot\! \sB(\bv \!-\! \bvp) , 
\label{eq:def_b_naive}
\\
\sD[F](\bv) & {} = m \!\!\int\!\! \rd \bvp \, \sB(\bv \!-\! \bvp) F(\bvp) .
\label{eq:def_D_naive}
\end{align}
\label{eq:def_b_D_naive}\end{subequations}
As already put forward in~\cite{Fu+2025},
by direct identification,
we can then write the standard Langevin process~\citep{Risken1989},
with It\^{o} formulation, as
\begin{equation}
\rd \bv = \bb[F](\bv) \, \rd t + \sD^{1/2} [F](\bv) \, \rd \bW_t .
\end{equation}
This equation describes the diffusive motion
of a particle embedded in a smooth background $F$.

Substituting ${ F_N }$ (see Eq.~\ref{def_F_N} in the main text) for $F$ yields a system of coupled Langevin equations describing the simultaneous evolution of all particles $i$.
It reads
\begin{align}
\rd \bv_{i} = {} & \bb[F_N](\bv_i) \rd t + \sD^{1/2}[F_N](\bv_i) \, \rd \bW_{i,t} .
\label{eq:Langevin_naive_unsoftened}
\end{align}
where the $\{ \mathrm{d} \mathbf{W}_{i,t} \}_{1 \leq i \leq N}$ are mutually independent $d$-dimensional Wiener processes.
Introducing the relative velocity ${ \bu_{ij} \!=\! \bv_{i} \!-\! \bv_{j} }$
and its norm ${ u_{ij} \!=\! | \bu_{ij} | }$, the drift and diffusion terms read
\begin{subequations}
\begin{align}
\bb[F_N](\bv_i) & {} = m^2 \sum_j \p_{\bv_i} \!\cdot\! \sB(\bu_{ij}) , 
\label{eq:b_naive_i}
\\
\sD[F_N](\bv_i) & {} = m^2 \sum_j \sB(\bu_{ij}) .
\label{eq:D_naive_i}
\end{align}
\label{eq:b_D_naive_i}\end{subequations}
Once again, we stress that Eq.~\eqref{eq:Langevin_naive_unsoftened}
does not conserve the total (kinetic) energy nor the total momentum.
Here, we introduce the square root ${ \sD^{1/2} }$, defined as ${ \sD \!=\! \sD^{1/2} \, [\sD^{1/2}]^{\rT} }$. We highlight that $\sD^{1/2}$ is not unique. Changing the square root matrix $\sD^{1/2}$ modifies exact particle paths for a fixed noise realization, but leaves the statistics of the Langevin process unchanged.

As pointed by~\cite{Fu+2025}, in this form,
Eq.~\eqref{eq:Langevin_naive_unsoftened}
cannot be easily simulated,
because of the ${ 1/u }$ divergence
of the collision kernel in Eq.~\eqref{eq:def_sB_generic}.
To circumvent this numerical divergence,
we introduce softening in the collision kernel.
For ${ d \!=\! 2 }$, this amounts to replacing Eq.~\eqref{eq:def_sB_generic} with 
\begin{equation}
f_{\eps}(u) = \kappa \, / (u \!+\! \eps)^3 .
\label{eq:def_fu_with_eps}
\end{equation}
Fortunately, adding this softening
does not affect the fluctuation-dissipation relation, ${\bb = \p_{\bv} \sB}$.
Using the expression for $\sB$ in Eq.~\eqref{eq:def_sB_generic},
the drift and diffusion terms in Eq.~\eqref{eq:def_b_D_naive} are replaced by their softened counterparts
\begin{subequations}
\begin{align}
\bb[F_N] (\bv_{i}) {} & = -m^2 \sum_{j \neq i} f_{\eps}(u_{ij}) \, \bu_{ij} .
\label{eq:coeffs_Langevin_Drift}
\\
\sD[F_N] (\bv_{i}) {} & = m^2 \sum_{j \neq i} f_{\eps}(u_{ij}) \, \sP(\bu_{ij}) ,
\label{eq:coeffs_Langevin_Diff}
\end{align}
\label{eq:coeffs_Langevin}\end{subequations}
The Stratonovich interpretation of the softened version of Eq.~\eqref{eq:Langevin_naive_unsoftened} reads
\begin{equation}
\rd \bv_{i} = \sD^{1/2}[F_N](\bv_i) \circ \rd \bW_{i,t} .
\label{eq:Langevin_naive_with_eps_Strato}
\end{equation}

\subsection{Integration scheme}
\label{app:Euler_Maruyama_integration}

In order to integrate Eq.~\eqref{eq:Langevin_naive_unsoftened},
we use the standard explicit Euler--Maruyama scheme~\citep[see, e.g.\@,][]{Kloeden+2011},
that is directly compatible with the It\^{o} interpretation.\footnote{In our numerical experiments, integrating Eq.~\eqref{eq:Langevin_naive_with_eps_Strato} using the implicit midpoint method, compatible with the Stratonovich interpretation, led to numerical instabilities on long times, even in the presence of softening (Eq.~\ref{eq:def_fu_with_eps}).}
Given a timestep $\DT$,
the evolution of the velocity $\bv_{i}$ of particle $i$
is governed by
\begin{equation}
\bv_{i} (t \!+\! \DT) = \bv_{i} (t) + \Delta \bv_{i} .
\label{eq:update_Langevin}
\end{equation}
Here, the velocity increments, ${ \Delta \bv_{i} }$, read
\begin{equation}
\Delta \bv_{i} = \bb (\bv_{i}) \, \DT + \sD^{1/2} (\bv_{i}) \, \DW_{i} ,
\label{eq:increment_Langevin}
\end{equation}
with the (independent) Wiener increments ${ \DW_{i} \!\sim\! \mN (0 , \DT \, \sI) }$.
In practice, for ${ d \!=\! 2 }$,
the computation of a ``square root'', $\sD^{1/2}$,
can be made explicitly
without resorting to any generic linear algebra solvers.

The Euler--Maruyama scheme from Eqs.~\eqref{eq:update_Langevin} and~\eqref{eq:increment_Langevin} has a weak convergence in time of first-order~\citep{Kloeden+2011}.

\subsection{Numerical parameters}
\label{app:parameters_naive}

In practice, to obtain Fig.~\ref{fig:Langevin_naive_contours},
we performed 10\,000 independent realizations
of Eq.~\eqref{eq:Langevin_naive_with_eps_Strato} using ${ N \!=\! 100 }$.
We set ${ \DT \!=\! 1 \!\times\! 10^{-5} \Tdyn }$
and set the softening to ${ \eps \!=\! \EPS \!=\! 0.01 \, v_{0} }$.
We performed a total of 1\,000 bootstrap resamplings
to estimate the uncertainties.
We checked that small variations in our choice of $\DT$ or $\EPS$
did not affect the results presented in Fig.~\ref{fig:Langevin_naive_contours}.
To ensure numerical stability, the maximum velocity increment during a close encounter must remain smaller than the softening scale $\epsilon$. This imposes the scaling condition ${ N^2 \Delta t \ll (\epsilon/v_0)^3 }$  on the time step.
This is safely satisfied by our chosen parameters. Finally, $\epsilon$ must be kept small compared to $v_0$ to mitigate any unphysical pollution of the macroscopic relaxation.

\section{Langevin-EP}
\label{app:EP_Langevin_simulations}

In this Appendix, we detail our implementation
of \LangevinEP\@ (see Eq.~\ref{eq:EP_Langevin} in the main text).
All the elements presented in this Appendix
are implemented in an efficient \texttt{julia} code,
which is publicly distributed~\cite{github}.
Additionally, we reproduce the analytical proofs from~\cite{Fu+2025} for the various properties of \LangevinEP discussed in the main text.

\subsection{Semi-implicit scheme}
\label{app:Semi-implicit_EP_Langevin}

We start with the \LangevinEP\@ process of Eq.~\eqref{eq:EP_Langevin} and insert the explicit collision kernel from Eq.~\eqref{eq:def_sB_generic}.
We get
\begin{equation}
\rd \bv_{i} \!=\! \sum_{j \neq i} \sqrt{m^2 \kappa \frac{1}{u_{ij}^3} \sP(\bu_{ij}) } \circ \rd \bW_{ij,t} ,
\label{eq:Langevin_EP_full_expression}
\end{equation}
with the projector, $\sP$, defined in Eq.~\eqref{eq:def_sP}.
To extract the projector from the square root, we express it via the unit vector ${ \hbu \!=\! \bu / u }$ such that ${ \sP(\bu) \!=\! u^2 \sP(\hbu) }$.
Applying the idempotence property, ${ \sP(\hbu)^2 \!=\! \sP(\hbu) }$,
and reverting to the unnormalized projector ${ \sP(\bu) }$, the differential equation simplifies to
\begin{align} 
\rd \bv_{i} & {} = \sqrt{m^2 \kappa} \, \sum_{j \neq i} \frac{1}{u_{ij}^{5/2}} \sP ( \bu_{ij} ) \circ \rd \bW_{ij,t}.
\end{align}

Numerically, we discretize time with a step $\Delta t$ and denote by $\bv_i^k$ the velocity of particle $i$ at time ${t_k \!=\! k \Delta t}$. Defining the velocity increment as ${\Delta \bv_i^k \!=\! \bv_i^{k+1} \!-\! \bv_i^k}$ and the midpoint velocity as ${\bv_i^{k+1/2} \!=\! (\bv_i^k \!+\! \bv_i^{k+1})/2}$, this leads to the implicit midpoint integration scheme,
\begin{equation}
\Delta \bv_i^k = \sum_{j \neq i} g \big(u^{k+1/2}_{ij} \big) \, \sP \big(\bu_{ij}^{k+1/2} \big) \DW_{ij,k} ,
\label{eq_app:Langevin_implicit}
\end{equation}
where ${\bu_{ij}^{k+1/2} \!=\! \bv_i^{k+1/2} \!-\! \bv_j^{k+1/2}}$ is the relative midpoint velocity and ${ u_{ij}^{k+1/2} = |\bu_{ij}^{k+1/2}|}$. The Wiener increments are anti-symmetric (see Eq.~\ref{eq:Noise_antisymmetry} in the main text) and, for notational convenience, we introduced the function
\begin{equation}
g(u) := \sqrt{m^2 \kappa} \frac{1}{u^{5/2}} .
\label{eq:def_g}
\end{equation} 

Due to its implicit nature, simulating this scheme requires an iterative approach, such as a fixed-point iteration.
However, as discussed in~\cite{Fu+2025},
because of the singularity in ${ g(u) }$, the fixed-point iteration search
can fail to converge.
To circumvent this issue,
\cite{Fu+2025} put forward a modified version of the implicit method. By writing the projector, $\sP$, as a two-variable function ${\sP(\bv, \bvp) \!=\! (\bv \!\cdot \bvp) \sI \!-\! \bv \! \otimes \bvp }$, this reads
\begin{equation}
\Delta \bv_i^k \!=\! \sum_{j \neq i} \! g (u^{k}_{ij} ) \sP \big( \bu_{ij}^{k} , \bu_{ij}^{k + 1/2} \big) \DW_{ij,k} ,
\label{eq_app:Langevin_semi_implicit}
\end{equation}
With this semi-implicit method, \cite{Fu+2025} makes the RHS of Eq.~\eqref{eq_app:Langevin_implicit} linear in $\bu_{ij}^{k+1/2}$. It can then be solved using a linear solver, avoiding potential convergence issues of an iterative method.
We stress that here no softening on $g$ (Eq.~\ref{eq:def_g}) is needed
for \LangevinEP\@ (Eq.~\ref{eq:EP_Langevin}).
In future work, following~\cite{Hong+2022}, we will investigate the design of higher-order integration schemes based on geometric methods.

\subsection{Matrix rewriting of the semi-implicit scheme}
\label{subsec:Cayley_Matrix}

In this section, we explicitly solve $\bu_{ij}^{k+1/2}$ in the semi-implicit method.
We start from the semi-implicit relation given by Eq.~\eqref{eq_app:Langevin_semi_implicit}.
It takes the explicit form
\begin{equation}
\Delta \bv^k_i = \sum_{j \neq i} g (u^{k}_{ij} ) \big[ \big( \bu_{ij}^{k} \!\cdot\! \bu_{ij}^{k + 1/2} \big) \sI \!-\! \bu_{ij}^{k} \!\otimes\! \bu_{ij}^{k+1/2} \big] \DW_{ij,k} ,
\label{eq:cpy_Cayley_start}
\end{equation}
In this expression, we recall that
\begin{subequations}
\begin{align}
\Delta \bv_{i}^{k} {} & = \bv^{k+1}_{i} - \bv^{k}_{i} ,
\label{eq:def_diff_Deltabv}
\\
\bu_{ij}^{k} {} & = \bv_{i}^{k} - \bv_{j}^{k} ,
\label{eq:def_diff_bu}
\\
\bu^{k+1/2}_{ij} {} & = \half [\bu^{k+1}_{ij} \!+\! \bu^{k}_{ij}] .
\label{eq:recall_diff}
\end{align}
\label{eq:def_diff_scheme}\end{subequations}
We then introduce the shortened notation
\begin{subequations}
\begin{align}
\bDelta_{i} {} & = \bv^{k+1}_{i} - \bv^{k}_{i} ,
\label{eq:def_bDelta}
\\
\bSigma_{i} {} & = \bv^{k+1}_{i} + \bv^{k}_{i} .
\label{eq:def_bSigma}
\end{align}
\label{eq:def_bDelta_bSigma}\end{subequations}
This allows us to rewrite Eq.~\eqref{eq:cpy_Cayley_start} as
\begin{align}
\bDelta_{i} = {} & \sum_{j} \big[ 1 \!-\! \delta_{i}^{j} \big] \half \, g_{ij} \big\{ \big[ \big\{ \bu_{ij} \!\cdot\! \big( \bSigma_{i} \!-\! \bSigma_{j} \big) \big\} \, \sI 
\nonumber
\\
- {} & \bu_{ij} \!\otimes\! \big( \bSigma_{i} \!-\! \bSigma_{j} \big) \big] \big\} \bW_{ij} ,
\label{eq:rwt_Cayley_1}
\end{align}
where we introduced the shortened notations ${ g_{ij} \!=\! g (u_{ij}^{k}) }$,
${ \bu_{ij} \!=\! \bu_{ij}^{k} }$,
and ${ \bW_{ij} \!=\! \DW_{ij, k} }$.
In Eq.~\eqref{eq:rwt_Cayley_1},
the factor ${ [1 \!-\! \delta_{i}^{j}] }$ is here to emphasize that the sum is only to be made for ${ j \!\neq\! i }$.

Finally, we introduce the index ${ [i,\alpha] }$,
with ${ 1 \!\leq\! \alpha \!\leq\! d }$
associated with the dimension.
Our goal is to obtain a matrix relation of the form
\begin{equation}
\Delta_{[i,\alpha]} = \sum_{j,\beta} Q_{[i,\alpha],[j,\beta]} \Sigma_{[j,\beta]} ,
\label{eq:intro_Q}
\end{equation}
where the matrix $\sQ$ only depends on ${ \{ \bv_{i}^{k} \}_{i,k} }$,
i.e.\ the coordinates at time $k$.
Following this relation, we have
\begin{equation}
Q_{[i,\alpha][j,\beta]} = \frac{\delta \Delta_{[i,\alpha]}}{\delta \Sigma_{[j,\beta]}} ,
\label{eq:Q_from_FuncDer}
\end{equation}
with the usual fundamental relation
\begin{equation}
\frac{\delta \Sigma_{[i,\alpha]}}{\delta \Sigma_{[j,\beta]}} = \delta_{i}^{j} \delta_{\alpha}^{\beta} .
\label{eq:fund_FuncDer}
\end{equation}

Armed with these notations, we rewrite Eq.~\eqref{eq:rwt_Cayley_1} as
\begin{align}
\Delta_{[i,\alpha]} = & {} \sum_{j , \beta} \half \big( 1 \!-\! \delta_{i}^{j} \big) g_{ij} \big\{ \big[ \big\{ \bu_{ij} \!\cdot\! \big( \bSigma_{i} \!-\! \bSigma_{j} \big) \big\} \, \sI
\nonumber
\\
- {} & \bu_{ij} \!\otimes\! \big( \bSigma_{i} \!-\! \bSigma_{j} \big) \big] \big\}_{\alpha \beta} W_{ij}^{\beta}
\nonumber
\\
= & {} \sum_{j,\beta} \half \big( 1 \!-\! \delta_{i}^{j} \big) g_{ij} \big\{ \delta_{\alpha}^{\beta} \sum_{\gamma} \big[ u_{ij}^{\gamma} \big( \Sigma_{[i,\gamma]} \!-\! \Sigma_{[j,\gamma]} \big) \big]
\nonumber
\\
- {} & u_{ij}^{\alpha} \big( \Sigma_{[i,\beta]} \!-\! \Sigma_{[j,\beta]} \big) \big\} W_{ij}^{\beta}
\nonumber
\\
= & {} \sum_{j,\beta,\gamma} \half \big( 1 \!-\! \delta_{i}^{j} \big) g_{ij} W^{\beta}_{ij} \big\{ \delta_{\alpha}^{\beta} \big[ u_{ij}^{\gamma} \big( \Sigma_{[i,\gamma]} \!-\! \Sigma_{[j,\gamma]} \big) \big]
\nonumber
\\
 - & {} \delta_{\alpha}^{\gamma} u_{ij}^{\gamma} \big( \Sigma_{[i,\beta]} \!-\! \Sigma_{[j,\beta]} \big) \big\} .
\label{eq:rwt_Cayley_2}
\end{align}
Making the change of labelling ${ \beta \!\leftrightarrow\! \gamma }$ in the second term,
we can rewrite this equation as
\begin{align}
\Delta_{[i,\alpha]} = \sum_{j , \beta , \gamma} {} & \half \big( 1 \!-\! \delta_{i}^{j} \big) g_{ij} \delta_{\alpha}^{\beta} \, \big( \Sigma_{[i,\gamma]} \!-\! \Sigma_{[j,\gamma]} \big)
\nonumber
\\
\times {} & \big( W_{ij}^{\beta} u_{ij}^{\gamma} \!-\! W_{ij}^{\gamma} u_{ij}^{\beta} \big) .
\label{eq:rwt_Cayley_3}
\end{align}
Executing the sum over $\beta$, and relabelling $j$ as $l$,
this expression becomes
\begin{align}
\Delta_{[i,\alpha]} = \sum_{l , \gamma} {} & \half \big( 1 \!-\! \delta_{i}^{l} \big) g_{il} \big( \Sigma_{[i,\gamma]} \!-\! \Sigma_{[l,\gamma]} \big)
\big( W_{il}^{\alpha} u_{il}^{\gamma} \!-\! W_{il}^{\gamma} u_{il}^{\alpha} \big).
\label{eq:rwt_Cayley_4}
\end{align}
We may now apply the relation from Eq.~\eqref{eq:Q_from_FuncDer}.
Differentiating Eq.~\eqref{eq:rwt_Cayley_4}
w.r.t.\ $\Sigma_{[j,\beta]}$, we find
\begin{align}
Q_{[i,\alpha][j,\beta]} {} & \!=\! \sum_{l,\gamma} \half \big( 1 \!-\! \delta_{i}^{l} \big) g_{il} \big( \delta_{i}^{j} \delta_{\gamma}^{\beta} \!-\! \delta_{l}^{j} \delta_{\gamma}^{\beta} \big) \big( W_{il}^{\alpha} u_{il}^{\gamma} \!-\! W_{il}^{\gamma} u_{il}^{\alpha} \big)
\nonumber
\\
{} & = \sum_{l} \half \big( 1 \!-\! \delta_{i}^{l} \big) g_{il} \big( \delta_{i}^{j} \!-\! \delta_{l}^{j} \big) \big( W_{il}^{\alpha} u_{il}^{\beta} \!-\! W_{il}^{\beta} u_{il}^{\alpha} \big) .
\label{eq:first_calc_Q}
\end{align}

We now consider in turn the cases ${ i \!=\! j }$
and ${ i \!\neq\! j }$.
For ${ i \!=\! j }$, Eq.~\eqref{eq:first_calc_Q} gives
\begin{align}
Q_{[i,\alpha][i,\beta]} {} & = \sum_{l} \half \big( 1 \!-\! \delta_{i}^{l} \big) g_{il} \, \big( W_{il}^{\alpha} u_{il}^{\beta} \!-\! W_{il}^{\beta} u_{il}^{\alpha} \big)
\nonumber
\\
{} & = \sum_{l \neq i} \half g_{il} \big( W_{il}^{\alpha} u_{il}^{\beta} \!-\! W_{il}^{\beta} u_{il}^{\alpha} \big) .
\label{eq:calc_Q_ii}
\end{align}
From this expression, we find the symmetries
\begin{equation}
Q_{[i,\alpha][i,\beta]} = - Q_{[i,\beta][i,\alpha]} .
\label{eq:sym_Q_ii}
\end{equation}
We now consider the case ${ i \!\neq\! j }$ in Eq.~\eqref{eq:first_calc_Q}.
We get
\begin{align}
Q_{[i,\alpha][j,\beta]} \big|_{j \neq i} {} & \!=\! - \sum_{l} \half \big( 1 \!-\! \delta_{i}^{l} \big) g_{il} \delta_{l}^{j} \, \big( W_{il}^{\alpha} u_{il}^{\beta} \!-\! W_{il}^{\beta} u_{il}^{\alpha} \big)
\nonumber
\\
{} & = - \half g_{ij} \big( W_{ij}^{\alpha} u_{ij}^{\beta} \!-\! W_{ij}^{\beta} u_{ij}^{\alpha} \big) .
\label{eq:calc_Q_ij}
\end{align}
From this relation, we find the symmetry
\begin{equation}
Q_{[j,\beta][i,\alpha]} \big|_{j \neq i} = - Q_{[i,\alpha][j,\beta]} ,
\label{eq:Sym_Q_ij}
\end{equation}
where we used the symmetries ${ g_{ji} \!=\! g_{ij} }$,
${ W_{ji}^{\alpha} \!=\! - W_{ij}^{\alpha} }$,
and ${ u_{ji}^{\alpha} \!=\! - u_{ij}^{\alpha} }$.
Gathering the symmetries from Eqs.~\eqref{eq:sym_Q_ii} and~\eqref{eq:Sym_Q_ij},
we find that the matrix $\sQ$ is skew-symmetric,
i.e.\ it satisfies
\begin{equation}
Q_{[j,\beta][i,\alpha]} = - Q_{[i,\alpha],[j,\beta]} .
\label{eq:skew_sym_Q}
\end{equation}
We note that this matrix also satisfies the generic relation
\begin{equation}
\forall i, j, \quad Q_{[i,\alpha],[j,\alpha]} = 0 .
\label{eq:vanish_semi_diag_Q}
\end{equation}
From this last relation, we recover that two-body deflections
cannot drive any relaxation for ${ d \!=\! 1 }$.

\subsection{Numerical integration}
\label{app:Numerical_integration}

For the numerical integration presented in Fig.~\ref{fig:Money_plot}, we proceed as follows.
We work in dimension $d$
and the state of the system is described by the collective velocity vector
${ \bV \!=\! (\bv_{1}, ..., \bv_{N}) \!\in\! \bbR^{dN} }$.

Given some timestep $\DT$,
for ${ 1 \!\leq\! i \!<\! j \!\leq\! N }$,
we generate the Wiener increments ${ \DW_{ij} \!\sim\! \mN (0 , \DT \, \sI) }$. Following Eq.~\eqref{eq:Noise_antisymmetry},
we then impose the antisymmetry condition
${ \DW_{ij} \!=\! - \DW_{ji} }$.
Then, following Eqs.~\eqref{eq:calc_Q_ii} and~\eqref{eq:calc_Q_ij},
we compute the matrix ${ \sQ \!=\! \sQ [\bV^{k}] }$,
enforcing explicitly its skew-symmetry (Eq.~\ref{eq:skew_sym_Q}),
in order to guarantee energy conservation.
Following Eq.~\eqref{eq:intro_Q},
the new state, $\bV^{k+1}$, is then given by the (semi-)implicit relation
\begin{equation}
\bV^{k+1} - \bV^{k} = \sQ \, \big[ \bV^{k+1} \!+\! \bV^{k} \big] .
\label{eq:implicit_relation_numeric}
\end{equation}
Equation~\eqref{eq:intro_Q} can be recast as
\begin{equation}
(\sI - \sQ) \, \bV^{k+1} = (\sI + \sQ) \, \bV^{k} ,
\label{eq:Cayley_scheme}
\end{equation}
with $\sI$ the identity matrix of size ${ dN \!\times\! dN }$.
To solve Eq.~\eqref{eq:Cayley_scheme}, we perform an LU factorization of ${ (\sI \!-\! \sQ) }$ at each timestep
and solve the associated linear system.

Before discussing further the convergence rate
of the scheme from Eq.~\eqref{eq:Cayley_scheme},
we first check that, indeed, our numerical implementation
conserves, up to round-off errors, the total momentum and energy,
as already put forward in~\citep{Fu+2025}.
In Appendix~\ref{app:dE_proof_Lang}, we provide the analytical derivation showing that these invariants are exactly conserved by \LangevinEP\@.

For the discretised scheme~\eqref{eq:Cayley_scheme}, the exact energy conservation stems directly from the properties of the Cayley transform. Indeed, Eq.~\eqref{eq:Cayley_scheme} can be rewritten as an explicit mapping ${ \bV^{k+1} \!=\! \sR \, \bV^{k} }$, where the transformation matrix is ${ \sR \!:=\! (\sI \!-\! \sQ)^{-1} (\sI \!+\! \sQ) }$. Because the matrix $\sQ$ is skew-symmetric,
the transpose of $\sR$ reads
\begin{align}
\sR^{\rT} = {} & (\sI + \sQ)^{\rT} \big( (\sI - \sQ)^{-1} \big)^{\rT}
\nonumber
\\
= {} & (\sI + \sQ^{\rT}) (\sI - \sQ^{\rT})^{-1}
\nonumber
\\
= {} & (\sI - \sQ) (\sI + \sQ)^{-1} .
\end{align}
Since the matrices ${ (\sI \!+\! \sQ) }$ and ${ (\sI \!-\! \sQ) }$ commute, we directly obtain ${ \sR^{\rT} \sR \!=\! \sI }$, meaning that the Cayley transform, $\sR$, is exactly orthogonal. Consequently, the discrete kinetic energy satisfies
\begin{align}
\| \bV^{k+1} \|^2 & {} = (\sR \bV^{k})^{\rT} (\sR \bV^{k})
\nonumber
\\
& {} = (\bV^{k})^{\rT} (\sR^{\rT} \sR) \bV^{k}
\nonumber
\\
& {} = (\bV^{k})^{\rT} \sI \bV^{k}
\nonumber
\\
& {} = \| \bV^{k} \|^2 ,
\end{align}
i.e.\ it is conserved during the integration.

\begin{figure}[htbp!]
\centering
\includegraphics[width=0.45 \textwidth]{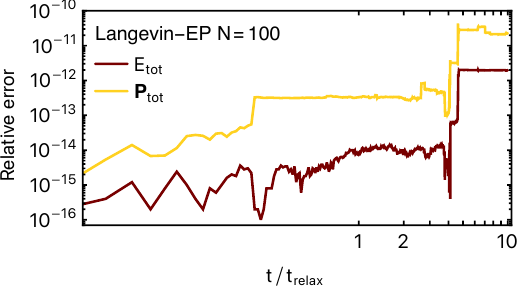}
\caption{Relative errors
in $\Etot$ and $\bPtot$ (w.r.t.\ their initial values)
for one realization of \LangevinEP\@ from Fig.~\ref{fig:Money_plot}.
As expected, the total energy and momentum are conserved,
up to round-off errors, except for a few abrupt jumps.
}
\label{fig:E_P_conservation}
\end{figure}

In Fig.~\ref{fig:E_P_conservation},
we explore the conservation of $\Etot$ and $\bPtot$
in one of the realization from Fig.~\ref{fig:Money_plot}.
In that figure, we note that $\Etot$ and $\bPtot$
are (very) well conserved for most of the relaxation.
In practice, sporadic spikes in errors can occur.
This likely results from the presence of very close deflections
in velocity space,
which probe the ${ 1/u }$ singularity of the collision kernel,
$\sB$ (Eq.~\ref{eq:def_sB_generic}).

\subsection{Numerical parameters}
\label{app:parameters_EP}

In practice, to obtain Fig.~\ref{fig:Money_plot},
we performed 10\,000 independent realizations
of Eq.~\eqref{eq:EP_Langevin} with ${d\!=\!2}$ and ${ N \!=\! 100 }$.
We set ${ \DT \!=\! 0.025 \, \Tdyn }$.
We performed a total of 1\,000 bootstrap resamplings
to estimate the uncertainties.
We checked that small variations in our choice of $\DT$
did not affect the results presented in Fig.~\ref{fig:Langevin_naive_contours}.

\subsection{Convergence rate}
\label{app:Langevin_convergence}

In order to prove that the semi-implicit method from Eq.~\eqref{eq_app:Langevin_semi_implicit}
converges to \LangevinEP\@ (Eq.~\ref{eq:EP_Langevin}),
we consider the numerical error as a function of the timestep $\DT$.
In the context of stochastic differential equations,
one must distinguish two types of convergence: strong and weak~\cite{Kloeden+2011}.

\begin{figure}[htbp!]
\centering
\includegraphics[width=0.45 \textwidth]{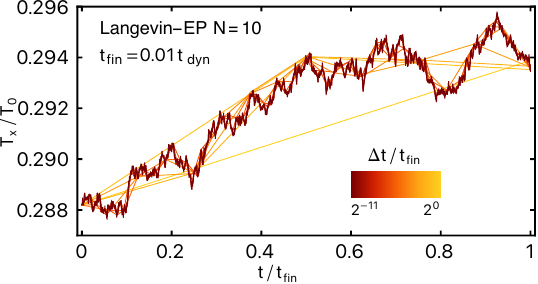}
\caption{Example of strong convergence on $T_{x}$ for one underlying \LangevinEP\@ with different time resolution levels ${ \DT \!=\! 2^{-\ell} }$. Visually, we can see that the convergence of each path when $\DT$ decreases.
}
\label{fig:strong_error_one_realization}
\end{figure}

A scheme is said to be of weak order $p_{\rw}$ (resp.\ strong order $p_{\rs}$)
if
\begin{subequations}
\begin{align}
| \langle \bVnum (T) \rangle - \langle \bVexact (T) \rangle | {} & \leq C \, \DT^{p_{\rw}} ,
\label{eq:def_pw}
\\
\big\langle | \bVnum (T) \!-\! \bVexact (T) |^2 \big\rangle^{1/2} {} & \leq C \, \DT^{p_{\rs}} .
\label{eq:def_ps}
\end{align}
\label{eq:def_pw_ps}\end{subequations}
With this definition, the weak error measures the average error between the final exact solution, ${ \bVexact (T) }$, and the simulated value, ${ \bVnum (T) }$:
it shows the systemic bias.
On the contrary, the strong convergence assesses the path-wise accuracy of the numerical scheme w.r.t.\ the exact solution.
As such, the strong error measures the deviation between a numerical trajectory and the exact trajectory of the system under the same realization of the noise, as illustrated in Fig.~\ref{fig:strong_error_one_realization}.
As demonstrated by~\cite{Fu+2025}, this numerical scheme yields a weak convergence of order ${ p_{\rw} \!=\! 1 }$ and a strong convergence of order ${ p_{\rs} \!=\! 1/2 }$. In the following, we aim to numerically reproduce and verify these theoretical rates for our specific setup.

Since the exact analytical solution, ${ \bVexact (t) }$, is unknown for \LangevinEP\@,
we adopt the same method as described in~\cite{Fu+2025}.
We consider a fixed final time $T$.
We define a series of time resolutions,
${ \DT_{\ell} \!=\! T / 2^{\ell} }$,
with ${ \ell \!\in\! \{0, 1, ..., L\} }$. The discrete time points at level $\ell$ are given by ${ t^{\ell}_k \!=\! k \, \DT_{\ell} }$.

In order to compute the strong error, it is crucial to build consistently the Wiener process across the different time levels $\DT_{\ell}$. To compare the trajectory computed with timestep $\DT_{\ell}$ to the one computed with $\DT_{\ell + 1}$, they must share the same underlying realization of the noise.
The Brownian increments at the level $\ell$, denoted ${ \DW^{\ell}_{k} }$, are constructed by summing the increments of the finer level ${ \ell \!+\! 1 }$ via
\begin{equation}
\DW^{\ell}_{k} = \DW^{\ell + 1}_{2k} + \DW^{\ell + 1}_{2k+1} .
\label{eq:Brownian_bridge}
\end{equation}
Since ${ \DW^{\ell+1} \!\sim\! \mN (0, \DT_{\ell+1} \sI) }$
and ${ \DT_{\ell} \!=\! 2 \, \DT_{\ell + 1} }$, this construction guarantees that ${ \DW^{\ell}_{k} \!\sim\! \mN (0, \DT_\ell \, \sI) }$,
while maintaining the path-wise correlation required for strong error analysis.

\begin{figure}[htbp!]
\centering
\includegraphics[width=0.45 \textwidth]{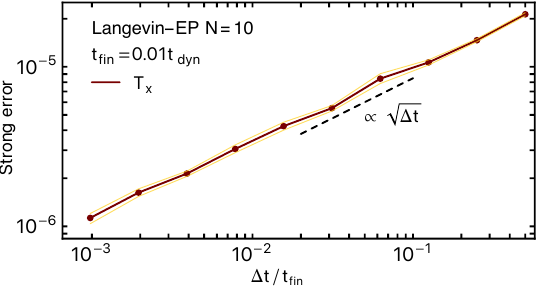}
\caption{Strong convergence error in the temperature $T_{x}$, as a function of the timestep $\DT$.
It clearly shows a strong rate convergence of order ${ 1/2 }$. Here, we sampled over ${ 2\,000 }$ realizations and, using bootstraps, we show with the yellow lines the 16\%--84\% spread away from the mean measurement.
}
\label{fig:strong_error_rate}
\end{figure}

Using this approach and following~\cite{Fu+2025}, we estimate the strong error by comparing the solution at level $\ell$ with the solution at the next finer level ${ \ell \!+\! 1 }$.
The relative strong error, ${ \eps_{\rs} (\DT_\ell) }$, is defined via 
\begin{equation}
\eps_{\rs} (\DT_\ell) = \big\langle | \bVnum (T; \DT_{\ell + 1}) \!-\! \bVnum (T; \DT_{\ell}) |^2 \big\rangle^{1/2} ,
\label{eq:strong_error_def}
\end{equation}
where the expectation ${ \langle \cdot \rangle }$ is the ensemble average over different Langevin realizations.
For the numerical demonstration presented in Fig.~\ref{fig:strong_error_rate},
we simulate a system of ${ N \!=\! 10 }$ particles
over the duration ${ \Tfin \!=\! 0.01 \, \Tdyn }$.
The ensemble average is computed over ${ 2\,000 }$ independent runs.
In Fig.~\ref{fig:strong_error_rate}, we show that the scheme from Eq.~\eqref{eq_app:Langevin_semi_implicit}
has a strong convergence rate of order ${ p_{\rs} \!=\! 1/2 }$.
This is consistent with the analytic result for this semi-implicit scheme,
as obtained in Appendix~{C3} of~\cite{Fu+2025}.
As noted in Appendix~{C4} of~\cite{Fu+2025}, if one takes too large a final time, $\Tfin$,
one can wrongly recover a convergence rate smaller than ${ 1/2 }$.

Additionally, we refrained from measuring the weak error, as extracting a statistically significant signal
demands a computationally prohibitive number of realizations
(see section~{III.B} in~\cite{Zhang+2020}).

\subsection{Stratonovich rewriting of Langevin-EP}
\label{app:Ito_to_Stratonovich}
To recover the Stratonovich formulation (Eq.~\ref{eq:EP_Langevin}) from the It\^{o} formulation (Eq.~\ref{eq:Langevin_EP_function_of_B}), we must carefully evaluate the noise-induced drift. This was already proven in Appendix~B3 of~\cite{Fu+2025}.
For a general system of It\^{o} stochastic differential equations of the form
\begin{equation}
\rd \bv_{i} = \bA_{i}^{\mathrm{Ito}} \, \rd t + \sigma_{i}(\bv_i) \, \rd \bW_{i,t} ,
\end{equation}
the equivalent Stratonovich equation is written as
\begin{equation}
\rd \bv_{i} = \bA_{i}^{\mathrm{Strato}} \, \rd t + \sigma_{i}(\bv_{i}) \circ \rd \bW_{i,t} ,
\label{eq:general_strato_form}
\end{equation}
where the modified Stratonovich drift is defined by \citep[see, e.g.,][]{Risken1989},
\begin{equation}
\bA_{i}^{\mathrm{Strato}} = \bA_{i}^{\mathrm{Ito}} - \bS_{i} ,
\label{eq:drift_conversion}
\end{equation}
with
\begin{equation}
    \bS_i = \half \big\langle \rd \big[\sigma_i(\bv_{i} )\big] \, \rd \bW_{i,t} \big\rangle .
    \label{eq:def_S_i}
\end{equation}
Here, $\bS_i$ represents the Stratonovich correction vector arising from the cross-variation between the diffusion matrix $\sigma_i$ and the Wiener process~\citep[see, e.g.,][]{Risken1989}.

In our specific case, the It\^{o} formulation (Eq.~\ref{eq:Langevin_EP_function_of_B}) gives the drift as
\begin{equation}
\bA_{i}^{\mathrm{Ito}} = \sum_{j \neq i} m^2 \p_{\bv_i} \!\cdot\! \sB(\bu_{ij}) ,
\label{eq:A_ito_def}
\end{equation}
and the corresponding diffusion term reads \begin{equation}
    \sum_{j \neq i} m \sB^{1/2}(\bu_{ji}) \,  \rd \bW_{ij,t} .
\end{equation}
However, because our system features pairwise interacting noise, the standard single-particle conversion formula must be adapted to account for the antisymmetry of the Wiener processes.

To compute the correction $\bS_i$, we must evaluate the differential of the noise coefficient ${ \rd \sB^{1/2}(\bu_{ij}) }$.
By the chain rule, this depends on the relative velocity differential ${ \rd\bu_{ij} \!=\! \rd\bv_i \!-\! \rd\bv_j }$. Retaining only the stochastic terms that contribute to the non-zero cross-variation
of order ${ \mO (\rd t) }$, we have
\begin{align}
\rd \bu_{ij} = {} & m \sB^{1/2}(\bu_{ij}) \rd \bW_{ij,t} \!-\! m \sB^{1/2}(\bu_{ji}) \rd \bW_{ji,t}
\nonumber
\\
+ \, {} & \mathcal{O}(\rd t) .
\end{align}
Because the pairwise noise is antisymmetric (${ \rd \bW_{ji,t} \!=\! -\rd \bW_{ij,t} }$) and the collision tensor is an even function (${ \sB^{1/2}(\bu_{ji}) \!=\! \sB^{1/2}(\bu_{ij}) }$), the two terms sum perfectly. This yields
\begin{equation}
\rd \bu_{ij} = 2m \sB^{1/2}(\bu_{ij}) \, \rd \bW_{ij,t} .
\label{eq:relative_vel_diff}
\end{equation}

From~\eqref{eq:def_S_i} and recalling that the Wiener processes here verify ${ \langle \rd \bW_{ij, t} \!\otimes\! \rd \bW_{ij, t}\rangle \!=\! \sI \, \rd t }$, we obtain
\begin{align}
\bS_i & {} = \half \sum_{j \neq i} \big\langle m \, \rd \big[\sB^{1/2}(\bu_{ij} )\big] \, \rd \bW_{ij,t} \big\rangle
\nonumber
\\
& {} = \half \sum_{j \neq i} m \big\langle \big[ \rd \bu_{ij} \!\cdot\! \p_{\bv_i} \big] \sB^{1/2}(\bu_{ij}) \, \rd \bW_{ij,t} \big\rangle
\nonumber
\\
& {} = \sum_{j \neq i} m^2 \big[ \sB^{1/2} (\bu_{ij}) \p_{\bv_i} \big] \!\cdot\! \sB^{1/2} (\bu_{ij}) .
\label{eq:Si_calc}
\end{align}
Notice that the usual $1/2$ prefactor inherent to the general Stratonovich correction (see Eq. \ref{eq:def_S_i}) cancels out exactly with the factor $2$ coming from the relative velocity differential in Eq.~\eqref{eq:relative_vel_diff}.

Next, we analyze the initial It\^{o} drift term, $\bA_{i}^{\mathrm{Ito}}$, from Eq.~\eqref{eq:A_ito_def}, which involves the divergence of the tensor ${ \sB \!=\! \sB^{1/2}\sB^{1/2} }$. Using the product rule, this divergence expands as
\begin{align}
\p_{\bv_i} \!\cdot\! \sB(\bu_{ij}) = {} & \big[ \sB^{1/2}(\bu_{ij}) \p_{\bv_i} \big] \!\cdot\! \sB^{1/2}(\bu_{ij})
\nonumber
\\
+ \, {} & \sB^{1/2}(\bu_{ij}) \big[ \p_{\bv_i} \!\cdot\! \sB^{1/2}(\bu_{ij}) \big] .
\label{eq:identity_grad_B}
\end{align}
By construction, the collision tensor projects orthogonally to the relative velocity, meaning ${ \sB^{1/2}(\bu_{ij}) \, \bu_{ij} \!=\! \mathbf{0} }$. Furthermore, the vector produced by the divergence ${ \p_{\bv_i} \!\cdot\! \sB^{1/2}(\bu_{ij}) }$ is colinear to $\bu_{ij}$. Multiplying the matrix ${ \sB^{1/2}(\bu_{ij}) }$ by a vector colinear to $\bu_{ij}$ yields exactly zero.
Thus, the second term in Eq.~\eqref{eq:identity_grad_B} vanishes, leading to
\begin{equation}
\p_{\bv_i} \!\cdot\! \sB(\bu_{ij}) = \big[ \sB^{1/2}(\bu_{ij}) \p_{\bv_i} \big] \!\cdot\! \sB^{1/2}(\bu_{ij}) .
\label{eq:ito_drift_simplified}
\end{equation}

Finally, substituting Eqs.~\eqref{eq:A_ito_def}, \eqref{eq:Si_calc}, and \eqref{eq:ito_drift_simplified} into the conversion formula from Eq.~\ref{eq:drift_conversion},
we evaluate the new Stratonovich drift via
\begin{equation}
\bA_{i}^{\mathrm{Strato}} = \sum_{j \neq i} m^2 \bigg[ \p_{\bv_i} \!\cdot\! \sB(\bu_{ij}) - \big[ \sB^{1/2} (\bu_{ij}) \p_{\bv_i} \big] \!\cdot\! \sB^{1/2} (\bu_{ij}) \bigg] .
\end{equation}
The two terms inside the brackets cancel one another exactly, yielding
\begin{equation}
\bA_{i}^{\mathrm{Strato}} = \bZero .
\end{equation}
Hence, from Eq.~\eqref{eq:general_strato_form},
we exactly recover the Stratonovich formulation driven solely by the multiplicative noise, namely
\begin{equation}
\rd \bv_{i} = \sum_{j \neq i} m \, \sB^{1/2} (\bv_i \!-\! \bv_j) \circ \rd \bW_{ij, t} .
\end{equation}
This is precisely Eq.~\eqref{eq:EP_Langevin}.

\subsection{Energy conservation for Langevin-EP}
\label{app:dE_proof_Lang}

In this Appendix, following Appendix~B2 of~\cite{Fu+2025},
we prove that \LangevinEP\@ conserves exactly the kinetic energy.
We start by applying the It\^{o} lemma to
\begin{equation}
\Ekin = \sum_{i} |\bv_i|^2 .
\end{equation}
This gives
\begin{equation}
\rd \Ekin = \sum_{i} \rd ( | \bv_i |^2 ) = \sum_{i} \big( 2 \bv_i^{\rT} \rd \bv_i + \rd \bv_i^{\rT} \rd \bv_i \big).
\label{eq:calc_cons_E}
\end{equation}

Let us start with the term ${ \rd \bv_i^{\rT} \rd \bv_i }$ in Eq.~\eqref{eq:calc_cons_E}.
Here, only terms of order ${ \rd t }$,
i.e.\ ${ \rd \bW^{\rT} \rd \bW }$,
survive.
Since the Wiener processes are independent for distinct pairs of particles (see Eq.~\ref{eq:stat_noise_Rostoker} in the End Matter), this leads to
\begin{align}
\rd \bv_i^{\rT} \rd \bv_i & {}= \sum_{j \neq i} \Tr \bigg[ \sqrt{m^2} \, \big[ \sB^{1/2}_{ij} \big]^{\rT} \, \sqrt{m^2} \, \sB^{1/2}_{ij} \bigg] \rd t
\nonumber
\\
& {} = m^2 \sum_{j \neq i} \Tr \big[ \sB_{ij} \big] \rd t .
\label{eq:dE_dvdv_term}
\end{align}
where we shortened the notation with ${ \sB_{ij} \!=\! \sB (\bu_{ij}) }$
and similarly for $\sB^{1/2}$.

Let us now deal with the linear term, ${ 2 \bv_i^{\rT} \rd \bv_i }$,
in Eq.~\eqref{eq:calc_cons_E}.
Substituting ${ \rd \bv_i }$ with Eq.~\eqref{eq:Langevin_EP_function_of_B},
we obtain
\begin{equation}
\sum_{i} 2 \bv_i^{\rT} \rd \bv_i = \sum_{\substack{i,j \\ i \neq j}} 2 \bv_i^{\rT} \bigg[ \bb_{ij} \, \rd t + \sqrt{m^2} \, \sB^{1/2}_{ij} \, \rd \bW_{ij,t}\bigg] ,
\label{eq:linear_term_dE}
\end{equation}
with the notation ${ \bb_{ij} \!=\! \bb (\bu_{ij}) }$.
Using the antisymmetry of ${ \rd \bW_{ij,t} }$,
the noise term here can be rewritten as
\begin{align}
\sum_{\substack{i,j \\ i \neq j}} 2 \bv_i^{\rT} \bigg[ \sqrt{m^2} \, \sB^{1/2}_{ij} \, \rd \bW_{ij,t}\bigg] {} & = \sum_{\substack{i,j \\ i \neq j}} \bu_{ij}^{\rT} \bigg[\sqrt{m^2} \, \sB^{1/2}_{ij} \, \rd \bW_{ij,t} \bigg]
\nonumber
\\
{} & = 0 ,
\end{align}
where we used that $\sB$ is a projector,
so that ${ \bu_{ij}^{\rT} \, \sB^{1/2}_{ij} \!=\! 0 }$.

For the drift term in Eq.~\eqref{eq:linear_term_dE},
using the antisymmetry of the drift, ${ \bb(-\bu) \!=\! -\bb(\bu) }$, and the fluctuation-dissipation relation, ${ \bb(\bu) \!=\! m^2 \p_{\bu} \!\cdot\! \sB(\bu) }$, we obtain
\begin{align}
\sum_{\substack{i,j \\ i \neq j}} 2 \bv_i^{\rT} \bb_{ij} \rd t & = \sum_{\substack{i,j \\ i \neq j}} (\bv_i - \bv_j)^{\rT} \bb_{ij} \rd t
\nonumber
\\
& = \sum_{\substack{i,j \\ i \neq j}} m^2 \bu_{ij}^{\rT} \big( \p_{\bu_{ij}} \!\cdot\! \sB_{ij} \big) \rd t .
\label{eq:calc_drift_cons_E}
\end{align}
To evaluate this scalar product,
we use the product rule for the divergence of a matrix multiplied by a vector, namely
\begin{equation}
\p_{\bu} \!\cdot\! \big[ \sB(\bu) \, \bu \big] = \bu^{\rT} \big[ \p_{\bu} \!\cdot\! \sB(\bu) \big] + \Tr \big[ \sB(\bu) \, \p_{\bu} \bu \big] .
\end{equation}
Since ${ \sB(\bu) \, \bu \!=\! \bZero }$, the total divergence in the left-hand side is strictly zero. Furthermore, since the Jacobian ${ \p_{\bu} \bu }$ is simply the identity matrix $\sI$, we find the generic relation
\begin{equation}
0 = \bu^{\rT} \big[ \p_{\bu} \!\cdot\! \sB(\bu) \big] + \Tr \big[ \sB(\bu) \big] .
\end{equation}
Used in Eq.~\eqref{eq:calc_drift_cons_E},
this yields
\begin{equation}
\bu_{ij}^{\rT} \big[ \p_{\bu_{ij}} \!\cdot\! \sB_{ij} \big] = - \Tr \big[ \sB_{ij} \big] .
\label{eq:dE_vdv_term}
\end{equation}
Substituting this identity back into the drift contribution and combining it with the diffusion term from Eq.~\eqref{eq:dE_dvdv_term} finally gives
\begin{align}
\rd \Ekin {} & = \sum_{\substack{i,j \\ i \neq j}} m^2 \Big( \Tr \big[ \sB_{ij} \big] - \Tr \big[ \sB_{ij} \big] \Big) \rd t
\nonumber
\\
{} & = 0 .
\end{align}
We have therefore recovered the exact conservation of the total kinetic energy,
as already shown in~\cite{Fu+2025}.

\section{Landau equation}
\label{app:Landau}

In this Appendix, we detail our integration scheme
for the Landau equation (Eq.~\ref{eq:Landau}).
We implemented all these elements in an efficient \texttt{julia} code,
which is publicly distributed~\cite{github}.

We place ourselves in ${ d \!=\! 2 }$.
We discretise each velocity direction, ${ (v_{x},v_{y}) }$,
within the domain ${ -\VMAX \!\leq\! v_{x}, v_{y} \!\leq\! \VMAX }$
in $\NV$ bins of width ${ \DV \!=\! 2 \, \VMAX / \NV }$.
More precisely, each bin
is centered around the location
\begin{equation}
v_{x}^{k} = - \VMAX + (k \!-\! \half) \, \DV
\quad \text{with} \quad
1 \!\leq\! k \!\leq\! \NV ,
\label{eq:pos_bin_Landau}
\end{equation}
and similarly for ${ \{ v_{y}^{l} \}_{1 \leq l \leq \NV} }$.

To compute the flux in Eq.~\eqref{eq:Landau},
we proceed in three successive steps.
First, we estimate the first-order derivatives
of the \DF\@, via the stencils
\begin{subequations}
\begin{align}
\p_{v_{x}} F [v_{x}^{1} , v_{y}^{l}] {} & \!\simeq\! \tfrac{1}{\DV} \big( F [v_{x}^{2},v_{y}^{l}] \!-\! F [v_{x}^{1} , v_{y}^{l}] \big) ,
\label{eq:stencil_Landau_1st}
\\
\p_{v_{x}} F [v_{x}^{k} , v_{y}^{l}] {} & \!\simeq\! \tfrac{1}{2 \, \DV} \big( F [v_{x}^{k+1},v_{y}^{l}] \!-\! F [v_{x}^{k-1} , v_{y}^{l}] \big) ,
\label{eq:stencil_Landau_2nd}
\\
\p_{v_{x}} F [v_{x}^{\NV} , v_{y}^{l}] {} & \!\simeq\! \tfrac{1}{\DV} \big( F [v_{x}^{\NV},v_{y}^{l}] \!-\! F [v_{x}^{\NV-1} , v_{y}^{l}] \big) .
\label{eq:stencil_Landau_1st_again}
\end{align}
\label{eq:stencil_Landau}\end{subequations}
We proceed similarly for the derivatives ${ \p_{v_{y}} F }$.

\begin{figure}[htbp!]
\centering
\includegraphics[width=0.45\textwidth]{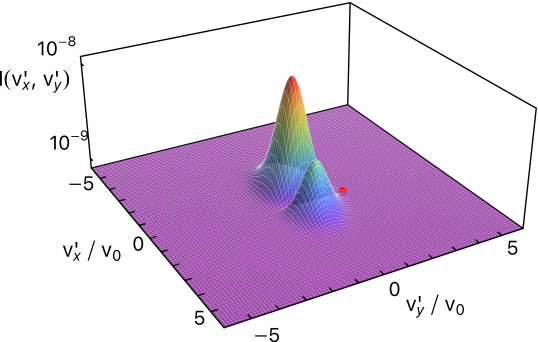}
\caption{Illustration of the dependence of the norm ${ I (\bvp) }$ of the integrand from Eq.~\eqref{eq:Landau}
as one varies $\bvp$, for some fixed ${ \bv / v_{0} \!=\! (0.52,1.72) }$.
In this figure, we used ${ \NV \!=\! 150}$ bins to discretize velocity space.
The integrand is a smooth function of $\bvp$
in particular for ${ \bvp \!\to\! \bv }$,
whose location is illustrated with the red dot.
}
\label{fig:integrand_Landau}
\end{figure}

Second, for each value of $\bv$,
we evaluate the integral over ${ \rd \bvp }$ in Eq.~\eqref{eq:Landau}
using a simple Riemann sum over the discrete bins.
Importantly, in order to improve the performance of the code,
we compute once and for all the interaction kernels,
${ \sB(\bv \!-\! \bvp) }$ (Eq.~\ref{eq:approx_sBNB}),
for all pairs of grid elements,
imposing ${ \sB (\bu \!=\! \bZero) \!=\! \sZero }$.
In Fig.~\ref{fig:integrand_Landau},
we illustrate the dependence of the norm, ${ I (\bvp) }$,
of the integrand from Eq.~\eqref{eq:Landau},
for some fixed $\bv$.
Reassuringly, we find that this integrand is smooth and shows
no sign of divergence for ${ \bvp \!\to\! \bv }$.
Once the integral from Eq.~\eqref{eq:Landau} estimated,
we use once again the formula from Eq.~\eqref{eq:stencil_Landau}
to compute the rates of change, ${ \p_{t} F }$,
for every bin.

Finally, once all the rates of changes estimated,
we integrate Eq.~\eqref{eq:Landau} forward in time
using the first-order Euler method.
Given some timestep, $\DT$, we perform
\begin{equation}
F (\bv , t \!+\! \DT) = F (\bv) + \DT \, \p_{t} F [\bv , t] .
\label{eq:integration_scheme_Landau}
\end{equation}

Our numerical integration of the Landau equation
involves a few control parameters,
in particular:
(i) $\VMAX$, the maximum extent of the considered region;
(ii) $\DV$, the width of the velocity bins;
(iii) $\DT$, the integration timestep.
In practice, given that Eq.~\eqref{eq:Landau}
involves singular functions (Eq.~\ref{eq:approx_sBNB}),
it is important to choose all these parameters carefully.
This is what we explore in the coming figures.

\begin{figure}[htbp!]
\centering
\includegraphics[width=0.45\textwidth]{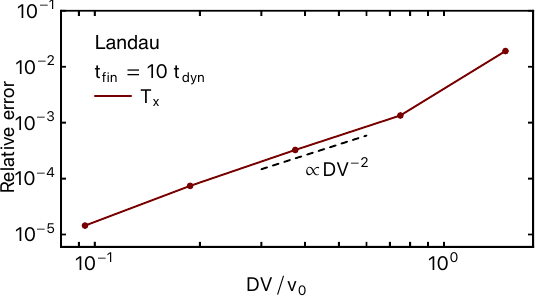}
\caption{Relative error in ${ T_{x} ( t \!=\! 10 \, \Tdyn) }$
as a function of the width of the velocity bins, $\DV$. Relative errors are computed w.r.t.\ the calculation with ${ \NV \!=\! 256 }$ (i.e.\ with ${ \DV / v_{0} \!=\! 4.7 \!\times\! 10^{-2} }$).
As expected, the finite difference scheme from Eq.~\eqref{eq:stencil_Landau} is second-order accurate.
}
\label{fig:cv_Landau_DV}
\end{figure}

\begin{figure}[htbp!]
\centering
\includegraphics[width=0.45\textwidth]{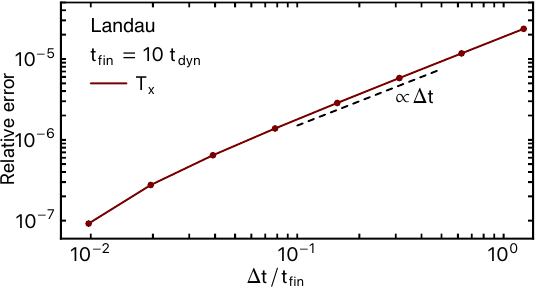}
\caption{Same as Fig.~\ref{fig:cv_Landau_DV},
but for the integration timestep, $\DT$. Relative errors are computed w.r.t.\ the calculation with ${ \DT / \Tdyn \!=\! 5 \!\times\! 10^{-3} }$.
As expected, the integration scheme from Eq.~\eqref{eq:integration_scheme_Landau} is first-order accurate.
}
\label{fig:cv_Landau_DT}
\end{figure}

For all the predictions presented in the main text,
we used ${ \VMAX \!=\! 6 \, V_0 }$,
${ \DV \!=\! 2 \, \VMAX / 128 }$ (i.e.\ ${ \NV \!=\! 128 }$);
${ \DT \!=\! 0.05 \, \Tdyn }$.
We now examine the dependence of our integration
w.r.t.\ variations in these parameters.
More precisely, we consider, in turn,
the convergence w.r.t.\
$\DV$ (Fig.~\ref{fig:cv_Landau_DV})
as well as w.r.t.\
$\DT$ (Fig.~\ref{fig:cv_Landau_DT}).

As anticipated, the spatial discretization is second-order accurate, with the numerical error scaling like ${ \mO (\DV^2) }$.
Using ${ \NV \!=\! 128 }$, the relative error drops to ${ \sim\! 10^{-5} }$. This confirms that the central difference stencil from Eq.~\eqref{eq:stencil_Landau} is robust enough to simulate the Landau equation over ${ 10 \, \Trelax }$. Similarly, in Fig.~\ref{fig:cv_Landau_DT},
we recover that the scheme is first-order accurate in $\DT$,
as expected from the standard explicit Euler method.

Because the Landau equation fundamentally behaves as a nonlinear diffusion equation, the explicit Euler method is subject to the standard stability constraint, ${ \DT \!\lesssim\! (\DV)^2 / \max(\sD) }$. Since the diffusion coefficient driving this slow relaxation is effectively small (${ \sD \!\propto\! m }$), the stability requirement is easily met,
and compatible with the timestep ${ \Delta t \!=\! 0.05 \, \Tdyn }$.

\section{From Langevin-EP to Large Deviations}
\label{app:Langevin_to_LD}

In this section, we analytically prove that \LangevinEP\@ accurately captures typical and large fluctuations. Specifically, we show that its corresponding \LD\@ Hamiltonian, ${ \mH[F, P] }$, is identical to the one previously derived for the Landau equation by~\cite{Feliachi+2021},
using a slow-fast approach.

Starting from Eq.~\eqref{eq:LDP_H}, the \LD\@ Hamiltonian can be calculated at the level of finite time increments through~\cite{Feliachi+2021}
\begin{equation}
\mH[P, F_N] = \lim_{\Delta \tau \to 0} \frac{1}{\Delta \tau} \frac{m}{\Mtot} \ln \big\langle \re^{\frac{1}{m} \langle\!\langle P, \Delta F_N(\tau) \rangle\!\rangle} \big\rangle \, \text{for} \, N \!\gg\! 1 ,
\end{equation}
where ${ \Delta F_N \!=\! F_N (\tau \!+\! \Delta \tau) \!-\! F_N(\tau) }$
and ${ \langle \cdot \rangle }$ denotes the conditional expectation over the realizations of the intrinsic stochastic noise, ${ \rd \bW }$, during the interval ${ \Delta \tau }$,
for a given instantaneous state ${ F_N(\tau) }$.
For a test function ${ \bv \!\mapsto\! P(\bv) }$,
we define the scalar product associated with a distribution function $F$ as
\begin{equation}
\langle\!\langle P, F \rangle\!\rangle := \!\! \int \!\! \rd \bv \, P(\bv) F(\bv) .
\end{equation}
In this framework, $P(\bv)$ acts as an observable that probes the state of the system, with the scalar product evaluating its macroscopic magnitude for a given distribution $F$~\cite{Bouchet2020}.
Another important point is that, here,
${ \Delta \tau }$ is the macroscopic time and differs from the previous Langevin time step $\Delta t$ by ${ \Delta \tau \!=\! (m/\Mtot) \Delta t }$.
We refer to~\cite{Feliachi+2024} and references therein for a more thorough presentation.
In terms of the macroscopic time, \LangevinEP\@ (see Eq.~\ref{eq:EP_Langevin} in the main text) is given by
\begin{equation}
\rd \bv_{i} = \sum_{j \neq i} \!\bigg[ m \Mtot \, \ba_{ij} \, \rd \tau \!+\! \sqrt{m \Mtot \sB_{ij} } \, \rd \bW_{ij,\tau} \!\bigg] ,
\end{equation}
where ${ \ba (\bu) \!=\! \p_{\bu} \!\cdot\! \sB (\bu) }$.
We also shortened the notation using
${ \ba_{ij} \!=\! \ba (u_{ij}) }$,
and similarly for ${ \sB_{ij} \!=\! \sB (\bu_{ij}) }$.
Using the definition of the empirical measure, we express the exponent ${ \langle\!\langle P, \, \Delta F_N(\tau) \rangle\!\rangle }$ in terms of the particles' increments ${ \Delta \bv_i }$ over a time interval ${ \Delta \tau }$.
We get
\begin{align}
\langle\!\langle P, & {} \, \Delta F_N(\tau) \rangle\!\rangle = m \sum_{i} \big[ P(\bv_i \!+\! \Delta \bv_i) - P(\bv_i) \big]
\nonumber
\\
= {} & m \sum_{i} \bigg[ \p_{\bv_i} P \!\cdot\! \Delta \bv_i + \half \Delta \bv_i^{\rT} \p_{\bv}^2 P(\bv_i) \Delta \bv_i \bigg] + o (\Delta \tau )
\nonumber
\\
= {} & m \sum_{\substack{i, j \\ i \neq j}} \p_{\bv_i} P \!\cdot\! \bigg[ m \Mtot \ba_{ij} \Delta \tau + \sqrt{ m \Mtot } \sB^{1/2}_{ij} \DW_{ij,\tau} \bigg]
\nonumber
\\
& {} + m \sum_{\substack{i, j \\ i\neq j}} \sum_{k \neq i} \half m \Mtot \big[ \sB^{1/2}_{ij} \DW_{ij,\tau} \big]^{\rT} \p_{\bv}^2 P(\bv_i)
\nonumber
\\
& {} \times \big[ \sB^{1/2}_{ik} \DW_{ik,\tau} \big] .
\label{eq:p_Dh}
\end{align}
where 
\begin{equation}
\big[ \p_{\bv}^2 P(\bv_i) \big]_{\alpha \beta} = \frac{\p^2 P}{\p v_{\alpha} \p v_{\beta}}(\bv_i)
\end{equation}
is the Hessian of $P$ w.r.t.\ $\bv_i$.
To obtain Eq.~\eqref{eq:p_Dh},
we perform a second-order Taylor expansion of ${ P(\bv_i \!+\! \Delta \bv_i) }$ around ${ \bv_i }$ using the rules of It\^{o} calculus.
We obtain three contributions: (i) the first-order gradient term associated with the deterministic drift $\ba$, (ii) the first-order gradient driven by the stochastic noise ${ \DW_{ij,\tau} }$, and (iii) the second-order Hessian term arising from the quadratic variation of the noise.
Crucially, because the increments ${ \DW_{ij,\tau} }$ have zero mean and are independent for distinct pairs, all cross-terms involving ${ \DW_{ij,\tau} }$ and ${ \DW_{ik,\tau} }$ (for ${ j \!\neq\! k }$) vanish upon taking the expectation. Furthermore, terms of higher order than ${ \mO(\Delta \tau) }$ vanish in the limit ${ \Delta \tau \!\to\! 0 }$.
Evaluating the logarithm of the expected exponential of Eq.~\eqref{eq:p_Dh}
yields the three distinct contributions
\begin{equation}
\ln \big\langle \re^{\frac{1}{m} \langle\!\langle P, \Delta F_N(\tau) \rangle\!\rangle} \big\rangle \!=\! \ln A_1 \!+\! \ln A_2 \!+\! \ln A_3 \!+\! o(\Delta \tau) .
\label{eq:split_Ai}
\end{equation}
Here, because cross-correlations between the linear and quadratic noise terms only contribute to ${\mO(\Delta \tau^2) }$, the expectation of the exponential factors into independent products up to ${o(\Delta \tau)}$.

Let us start by examining the first contribution in Eq.~\eqref{eq:split_Ai}.
From Eq.~\eqref{eq:p_Dh}, the expectation of the exponential associated with this first term reads
\begin{align}
A_1 & {} = \exp \bigg\{m \Mtot \sum_{\substack{i, j \\ i \neq j}} \p_{\bv_i} P \!\cdot\! \ba_{ij} \, \Delta \tau \bigg\}
\label{eq:calc_A1}
\\
& {} = \exp \bigg\{ \frac{\Mtot}{m} \!\!\int \!\! \rd \bv \rd \bvp \p_{\bv} P \!\cdot\! \ba(\bv \!-\! \bvp) F_N(\bv) F_N(\bvp) \Delta \tau \! \bigg\} .
\nonumber
\end{align}
Note that to pass from the discrete sum to the continuous integral over $F_N$,
we added the self-interaction term (${i\!=\!j}$). Since the drift vector $\ba$ is an odd function, this self-interaction term integrates to zero by symmetry, introducing no error.

For the second contribution, we have first to account for the antisymmetry property, ${ \DW_{ij,\tau} \!=\! - \DW_{ji,\tau} }$.
This implies
\begin{equation}
\sum_{\substack{i, j \\ i \neq j}} \! \big( \p_{\bv_i} P \big)^{\rT} \!\sB^{1/2}_{ij} \DW_{ij,\tau} \!=\! \sum_{\substack{i, j \\ i \neq j}} \! \half \bigg[\! \frac{\p P}{\p \bv_i} \!-\! \frac{\p P}{\p \bv_j} \!\bigg]^{\rT} \!\! \sB^{1/2}_{ij} \DW_{ij,\tau} .
\end{equation}
However, the increments here are still not independent, since we are including both $\DW_{ij,\tau}$ and $\DW_{ji,\tau}$. To make the variables truly independent and identically distributed, we restrict the sum to strictly unique pairs ${ (i \!<\! j) }$.
This is a crucial point of the calculation.
We find
\begin{equation}
\sum_{\substack{i, j \\ i \neq j}} \left(\p_{\bv_i} P\right)^{\rT} \! \sB^{1/2}_{ij} \DW_{ij,\tau} \!=\! \sum_{\substack{i,j \\ i < j}} \bigg[\! \frac{\p P}{\p \bv_i} \!-\! \frac{\p P}{\p \bv_j} \! \bigg]^{\rT} \! \sB^{1/2}_{ij} \DW_{ij,\tau} .
\end{equation}
Therefore, the second contribution can be rewritten as
\begin{align}
A_2
\!=\! {} & \bigg\langle\! \!\exp \!\bigg\{\!\sum_{\substack{i,j \\ i < j}}\! \bigg[\! \frac{\p P}{\p \bv_i} \!-\! \frac{\p P}{\p \bv_j} \!\bigg]^{\rT} \!\!\sqrt{m \Mtot} \, \sB^{1/2}_{ij} \DW_{ij, \tau} \!\bigg\} \!\bigg\rangle
\nonumber
\\
\!=\! {} & \bigg\langle\! \prod_{\substack{i,j \\ i < j}} \exp \!\bigg\{\! \sqrt{m\Mtot} \bigg[ \frac{\p P}{\p \bv_i} \!-\! \frac{\p P}{\p \bv_j} \bigg]^{\rT} \! \sB^{1/2}_{ij} \DW_{ij,\tau} \!\bigg\} \!\bigg\rangle .
\label{eq:compute_A2}
\end{align}
For ${ i \!<\! j }$, the ${ \DW_{ij,\tau} }$ are independent
from another. As a consequence, we can write
\begin{equation}
A_2 \!=\! \prod_{\substack{i,j \\ i < j}} \!\bigg\langle\! \!\exp \!\bigg\{\! \sqrt{m\Mtot} \bigg[\! \frac{\p P}{\p \bv_i} \!-\! \frac{\p P}{\p \bv_j} \!\bigg]^{\rT} \sB^{1/2}_{ij} \DW_{ij, \tau} \!\bigg\} \!\!\bigg\rangle .
\label{eq:second_term}
\end{equation}
For ${ i \!<\! j }$, one has ${ \DW_{ij,\tau} \!\sim\! \mN (0, \Delta \tau \, \sI) }$,
and for any vector $\bA$, we generically have
\begin{equation}
\big\langle \re^{\bA \cdot \DW} \big\rangle = \!\!\int\!\! \rd \bv \, \re^{\bA \cdot \bv \sqrt{\Delta \tau} } \re^{-v^2/2} = \re^{ |\bA |^2 \Delta \tau/2}.
\end{equation}
As a result, Eq.~\eqref{eq:second_term} becomes
\begin{align} 
A_2 = {} & \prod_{\substack{i,j \\ i < j}} \! \exp \!\bigg\{ \half \Mtot m \bigg[\! \frac{\p P}{\p \bv_i} \!-\! \frac{\p P}{\p \bv_j} \!\bigg]^{\rT}\!\!\sB_{ij} \bigg[\! \frac{\p P}{\p \bv_i} \!-\! \frac{\p P}{\p \bv_j} \!\bigg] \Delta \tau \!\bigg\}
\nonumber
\\
= {} & \exp \!\bigg\{\! \tfrac{1}{2} \Mtot m \sum_{\substack{i,j \\ i < j}} \!\bigg[\! \frac{\p P}{\p \bv_i} \!-\! \frac{\p P}{\p \bv_j} \!\bigg]^{\rT} \!\! \sB_{ij} \! \bigg[\! \frac{\p P}{\p \bv_i} \!-\! \frac{\p P}{\p \bv_j} \!\bigg] \, \Delta \tau \bigg\} .
\end{align}
To express the discrete sum as a continuous integral over $F_N$,
we sum over all ${ i \!\neq\! j }$.
This yields an additional factor of ${ 1/2 }$,
preventing double-counting.
We get
\begin{align}
A_2
\!=\! {} & \exp \!\bigg\{\! \tfrac{1}{4} \Mtot m \sum_{\substack{i,j \\ i \neq j}} \!\bigg[\! \frac{\p P}{\p \bv_i} \!-\! \frac{\p P}{\p \bv_j} \!\bigg]^{\rT} \!\sB_{ij} \bigg[\! \frac{\p P}{\p \bv_i} \!-\! \frac{\p P}{\p \bv_j} \!\bigg] \, \Delta \tau \!\bigg\} 
\nonumber
\\
\!=\! {} & \exp \bigg\{ \tfrac{1}{4} \frac{\Mtot}{m} \!\!\int\!\! \rd \bv \rd \bvp \, \bigg[\! \frac{\p P}{\p \bv} \!-\! \frac{\p P}{\p \bvp} \!\bigg]^{\rT}
\nonumber
\\
\times \, & {} \sB(\bv\!-\!\bvp) \bigg[\! \frac{\p P}{\p \bv} \!-\! \frac{\p P}{\p \bvp} \!\bigg] F_N(\bv) F_N(\bvp) \Delta \tau \bigg\} .
\label{eq:calc_A2}
\end{align}

For the last term in Eq.~\eqref{eq:p_Dh}, using the correlation between two Wiener processes (Eq.~\ref{eq:stat_noise_Rostoker}), we have
\begin{align}
A_3 = {} & \exp \bigg\{ \half m \Mtot \sum_{\substack{i, j \\ i \neq j}}\sum_{k \neq i} \p_{\bv}^2 P(\bv_i) \!:\! \big[ \sB^{1/2}_{ij} \sB^{1/2}_{ik} \big] \delta_{kj} \Delta \tau \bigg\}
\nonumber
\\ 
= {} & \exp \bigg\{ \half \Mtot m \sum_{\substack{i, j \\ i \neq j}} \p_{\bv}^2 P(\bv_i) \!:\! \sB_{ij} \, \Delta \tau \bigg\}
\nonumber
\\ 
= {} & \exp \bigg\{ \half \frac{\Mtot}{m} \!\!\int\!\! \rd \bv \rd \bvp \p_{\bv}^2 P(\bv) \!:\! \sB(\bv \!-\! \bvp)
\nonumber
\\
& {} \times F_N(\bv) F_N(\bvp) \Delta \tau \bigg\} . 
\label{eq:calc_A3_in_progress}
\end{align}
where the operator ``$:$'' denotes the contraction of two second order symmetric tensors,  ${ \sA \!:\! \sB \!=\! \Tr (\sA^{\rT} \sB) }$. 
Again, extending the integral to include the self-interaction term ${\bv \!=\! \bvp}$ introduces no error because ${\sB(\bZero) \!=\! \sZero }$.
We now assemble the full \LD\@ Hamiltonian for the \LangevinEP\@ dynamics. First, we take the logarithm of the product of the three contributions derived in Eqs.~\eqref{eq:calc_A1}, \eqref{eq:calc_A2}, and \eqref{eq:calc_A3_in_progress}. Next, we divide this result by ${\Delta \tau}$ and take the continuous-time limit ${\Delta \tau \to 0}$. Finally, taking the macroscopic limit ${N \to +\infty}$ allows us to safely replace the empirical measure $F_N$ with the mean distribution function $F$. This sequence of operations yields
\begin{align}
{} & \mHEP[F, P] = \!\!\int \!\! \rd \bv \rd \bvp \bigg\{ \ba(\bv \!-\! \bvp) \!\cdot\! \frac{\p P}{\p \bv} + \half \sB(\bv \!-\! \bvp) \!:\! \frac{\p^2 P}{\p \bv ^2}
\nonumber
\\
& {} + \tfrac{1}{4} \bigg[\! \frac{\p P}{\p \bv} \!-\! \frac{\p P}{\p \bvp} \!\bigg]^{\rT} \!\!\sB(\bv \!-\! \bvp) \bigg[\! \frac{\p P}{\p \bv} \!-\! \frac{\p P}{\p \bvp} \!\bigg] \!\bigg\} F(\bv) F(\bvp) .
\label{eq:LD_Hamiltonian}
\end{align}
The Hamiltonian can be decomposed as
\begin{equation}
\mHEP[F, P] = \mH^{(1)}[F, P] + \mH^{(2)}[F, P]
\label{eq:def_mH}
\end{equation}
where the first contribution is linear in $P$
\begin{align}
\mH^{(1)}[F, P] = {} & \half \!\!\int \!\! \rd \bv \rd \bvp \, P(\bv)
\label{eq:calc_H1}
\\
& {} \times \frac{\p}{\p \bv} \!\cdot\! \bigg\{\sB(\bv \!-\! \bvp) \bigg[ \frac{\p F}{\p \bv} F(\bvp) \!-\! \frac{\p F}{\p \bvp} F(\bv) \bigg] \bigg\} ,
\nonumber
\end{align}
and the second contribution is quadratic in $P$
\begin{align}
\mH^{(2)}[F, P] = {} & \half \!\!\int \!\! \rd \bv \rd \bvp \, \bigg[ \frac{\p P}{\p \bv} \bigg]^{\rT} \sB(\bv \!-\! \bvp) \bigg[ \frac{\p P}{\p \bv} \!-\! \frac{\p P}{\p \bvp} \bigg]
\nonumber
\\
\times \, & {} F(\bv) F(\bvp) .
\end{align}
This is the expression reproduced in the main text (Eq.~\ref{eq:mH_EP}).

The \LD\@ Hamiltonian matches with the one obtained in~\cite{Feliachi+2021}
from the slow-fast approach for the Landau equation (Eq.~\ref{eq:Landau}).
This result provides analytical confirmation of our central claim:
\LangevinEP\@ accurately predicts the full statistics of fluctuations.
As discussed in~\cite{Feliachi+2021}, the quadratic structure in $P$ of $\mHEP$
arises from the Gaussianity assumption underlying the \LangevinEP\@ dynamics~\eqref{eq:EP_Langevin}:
only the first two cumulants are needed.
When collective effects are taken into account, \cite{Feliachi+2022} shows that the \LD\@ Hamiltonian is no longer quadratic: all cumulants are needed.
Consequently, \LangevinEP\@ fails to capture the full fluctuation statistics when significant collective effects are present.
This is illustrated in Fig.~\ref{fig:Contours_G_10}.

We conclude this Appendix by performing the same calculation
for \LangevinNaive\@ (Eq.~\ref{eq:Langevin_naive}),
following~\cite{Feliachi+2021}.
The only difference with \LangevinEP\@
lies in the antisymmetry constraint on the noise ${ \rd \bW_{ij,\tau} }$.
Thus, the only difference in the derivation arises in the computation of $A_{2}$ in Eq~\eqref{eq:compute_A2}.
For \LangevinNaive\@ (Eq.~\ref{eq:Langevin_naive}),
we have
\begin{align}
A_2
{} & = \bigg\langle \exp \bigg\{ \sqrt{m \Mtot} \sum_{\substack{i, j \\ i \neq j}} \bigg[ \frac{\p P}{\p \bv_i} \bigg]^{\rT} \sB^{1/2}_{ij} \DW_{i,\tau} \bigg\} \bigg\rangle .
\label{eq:compute_A2_naive}
\end{align}
Evaluating this Gaussian expectation and combining it with the remaining contributions yields the \LD\@ Hamiltonian
for the \LangevinNaive\@ process, namely
\begin{align}
\mHNaive[F, P] = {} & \mH^{(1)}[F, P] 
\label{eq:calc_Hnaive}
\\
+ {} & \half \!\!\int \!\! \rd \bv \rd \bvp \left(\frac{\p P}{\p \bv}\right)^{\rT} \sB(\bv \!-\! \bvp) \frac{\p P}{\p \bv} F(\bv) F(\bvp).
\nonumber
\end{align}
As discussed in~\cite{Feliachi+2021},
this corresponds to the Hamiltonian that describes
$N$ particles diffusing with a mean field coupling.

\medskip
{\it Discussion} --- From the previous results,
another decomposition of the \LD\@ Hamiltonian for \LangevinEP\@ is given by
\begin{equation}
\mHEP[F, P] = \mHNaive[F, P] + \mHI[F, P],
\end{equation}
where $\mHNaive$ represents the contributions associated with the individual diffusion of particles due to their interaction with the mean field while
\begin{equation}
\mHI [F, P] \!=\! - \half \!\!\int \!\! \rd \bv \rd \bvp \left(\frac{\p P}{\p \bvp}\right)^{\rT} \!\! \sB(\bv \!-\! \bvp) \frac{\p P}{\p \bv} F(\bv) F(\bvp) ,
\end{equation}
represents the new contribution that accounts
for the correlations between particles
induced by momentum conservation (Eq.~\ref{eq:Noise_antisymmetry}).
Indeed, during a deflection between particles ${ (\bv, \bvp) }$,
a fluctuation ${ \delta \bv }$ in $\bv$ affects the particle with velocity $\bvp$ by ${ -\delta \bv }$. A binary deflection therefore produces simultaneous variations of ${ F(\bv) }$ and ${ F(\bvp) }$. Thus, since $P$ is the conjugate field of $F$, this leads to the cross term ${ \p_{\bv} P \, \p_{\bvp} P }$ in $\mH$.
As such, it is not surprising that the contribution $\mHI$ is absent from $\mHNaive$,
since it does not account for momentum conservation.
As a consequence, \LangevinNaive\@ does not predict the correct statistics of fluctuations, as observed in Fig.~\ref{fig:Langevin_naive_contours} of the main text.

\section{From Rostoker to Langevin-EP}
\label{app:Rostoker_to_Langevin}

In this Appendix, we explicitly compute the drift and diffusion terms of \LangevinEP\@,
following the approach of~\cite{Heyvaerts+2017} adapted here for a homogeneous, infinite system without collective effects.

\subsection{Drift term}
\label{app:Drift_computation}

The drift term ${ \bb(\bv_i, \bv_j) }$ in Eq.~\eqref{eq:def_b_Rostoker}
acting on particle $i$ due to deflections from particle $j$ over a time interval ${ \Tdyn \ll \Delta t \ll \Trelax }$ is defined as~\cite{Risken1989}
\begin{align} 
\bb(\bv_i, \, \bv_j) & {} = \lim_{\Delta t \to + \infty} \langle \Delta \bv_i^{j}\rangle_{\bT} / \Delta t
\nonumber
\\ 
& {} = \lim_{\Delta t \to + \infty} \, n \, \langle \delta \bv_i^{j }\rangle_{\bT} / \Delta t , \label{eq:def_b_calc_Rostoker} 
\end{align}
where the second equality follows from treating the $n$ individual velocity increments, ${ \{ \delta \bv_i^{j,(p)} \}_{1 \leq p \leq n} }$, as independent and identically distributed random variables. We recall that the total increment ${ \Delta \bv_i^{j} }$ depends only on the initial velocities $\bv_i$ and $\bv_j$ at time $t$, prior to the deflections. Because these unperturbed velocities are treated as constant over the deflection interval ${ \Delta t }$, every individual deflection $(p)$ depends on the same initial velocities $\bv_i$ and $\bv_j$. Therefore, to compute the expected microscopic increment ${ \langle \delta \bv_i^{j,(p)} \rangle_{\bT} }$, we simply average over the initial relative positions for each independent deflection $(p)$.

To streamline the notation, we introduce the sub-collision time
${ \delta t \!:=\! \Delta t / n }$,
the initial relative coordinates ${ \bT_{ij} \!:=\! \bT_{i} (0) \!-\! \bT_{j} (0) }$
and ${ \bv_{ij} \!:=\! \bv_i(0) \!-\! \bv_j(0) }$,
and the unperturbed phase frequency ${ \omega_{\bk} \!:=\! \bk \!\cdot\! \bv_{ij} }$.

We also define the perturbed phase ${ z_{ij}(t) \!:=\! \bk \!\cdot\! [\bT_i(t) \!-\! \bT_j(t)] }$.
At zeroth order, we have
\begin{equation}
z_{ij}(t) = \bk \!\cdot\! \bT_{ij} + \omega_{\bk} \, t .
\end{equation}
Evaluating the integral in Eq.~\eqref{eq:deltav_Rostoker} along this unperturbed linear trajectories yields
\begin{equation}
\delta \bv_{i}^{j} \!=\! - m \sum_{\bk} \ri \bk \, \psi_{\bk} \, \re^{\ri \bk \cdot \bT_{ij}} \, \frac{\re^{\ri \bk \cdot \bu_{ij} \delta t} \!-\! 1}{\ri \bk \!\cdot\! \bu_{ij}} ,
\label{eq:unperturbed_deltaV}
\end{equation}
with the notation ${ \bu_{ij} \!=\! \bv_{i} \!-\! \bv_{j} }$.
By averaging over the initial positions $\bT_i$ and $\bT_j$ directly, we end up with a null drift since only the ${ \bk \!=\! 0 }$ component survives the spatial average. Thus, we must compute the drift to second order in the interaction. From the definition of ${ z_{ij}(t) }$, we obtain the integral solution
\begin{align}
z_{ij}(t) = {} & \bk \!\cdot\! \bT_{ij} + \omega_{\bk} \, t
\nonumber
\\
& {} + \!\int_0^t \!\! \rd \tp \!\! \int_0^{\tp} \!\! \rd s \, \bk \!\cdot\! \big[ \dot{\bv}_i(s) - \dot{\bv}_j(s) \big] ,
\end{align}
where ${ \dot{\bv}_i (s) \!:=\! \p_s \bv_i (s) }$ is the acceleration of particle $i$.
Then, the exponential $\re^{\ri z_{ij}(t)}$ can be Taylor-expanded to first order in the acceleration. This gives
\begin{align}
\re^{\ri z_{ij}(t)} \simeq {} & \re^{\ri (\bk \cdot \bT_{ij} + \omega_{\bk} t)}
\label{eq:exp_expansion_drift}
\\
\times \, {} & \bigg\{ 1 + \ri \!\! \int_0^t \!\! \rd \tp \!\! \int_0^{\tp} \!\!\!\! \rd s \, \bk \!\cdot\! \big[ \dot{\bv}_{i} (s) \!-\! \dot{\bv}_{j} (s) \big] \bigg\} .
\nonumber
\end{align}
To compute these accelerations at leading order, we evaluate the interaction force along the unperturbed trajectories. From Eq.~\eqref{eq:unperturbed_deltaV}, we have
\begin{equation}
\dot{\bv}_i(s) = -m \sum_{\bkp} \ri \bkp \psi_{\bkp} \re^{\ri \bkp \cdot \bT_{ij}} \re^{\ri \omega_{\bkp} s} .
\end{equation} 
Then, Eq.~\eqref{eq:exp_expansion_drift} becomes
\begin{align}
& \re^{\ri z_{ij}(t)} = \re^{\ri \bk \cdot \bT_{ij} + \ri \omega_{\bk} t} \bigg\{ 1 \!+ \!\!\int_0^t \!\! \rd \tp \!\!\int_0^{\tp} \!\!\!\! \rd s
\label{eq:exp_expansion_drift_continued}
\\
& {} \times \ri \bk \!\cdot\! \bigg[ \!-\!m \sum_{\bkp} \psi_{\bkp} \ri \bkp \, \big( \re^{\ri \bkp \cdot \bT_{ij}(s)} \!-\! \re^{\ri \bkp \cdot \bT_{ji}(s)} \big) \bigg] \bigg\}
\nonumber
\\
& {} = \re^{\ri \bk \cdot \bT_{ij} + \ri \omega_{\bk} t} \bigg\{ 1 + Z_i^j - Z_j^i \bigg\} .
\nonumber
\end{align}
with
\begin{equation}
Z_i^j := \!\!\int_0^t \!\! \rd \tp \!\!\int_0^{\tp} \!\!\!\! \rd s \, \ri \bk \!\cdot\! \bigg[ \!-\!m \sum_{\bkp} \psi_{\bkp} \ri \bkp \re^{\ri \bkp \cdot \bT_{ij}(s)} \bigg] .
\end{equation}
When substituting Eq.~\eqref{eq:exp_expansion_drift_continued} into the velocity increment integral from Eq.~\eqref{eq:deltav_Rostoker}, the zeroth-order term (which corresponds to the ``$1$'' inside the brackets) averages to zero.
This leaves us with
\begin{align}
\delta \bv_i^{j} & {} = - \!\! \int_0^{\delta t} \!\! \rd t \, m \sum_{\bk} \ri \bk \, \re^{\ri z_{ij}(t)} \psi_{\bk}
\nonumber
\\
& {} = A_{i}^{j} + A_{j}^{i} ,
\label{eq:deltaV_A_ij}
\end{align}
where
\begin{equation}
A_{i}^{j} = - \!\! \int_0^{\delta t} \!\! \rd t \, m \sum_{\bk} \ri \bk \, \re^{\ri \bk \cdot \bT_{ij} + \ri \omega_{\bk} t} Z_i^j \psi_{\bk} ,
\end{equation}
corresponding to the contributions of ${ \dot{\bv}_i (s) }$.
Focusing on $A_{i}^{j}$, we have
\begin{align}
A_{i}^{j} {} & = -\ri m^2 \sum_{\bk, \bkp} \bk \, (\bk \!\cdot\! \bkp) \, \psi_{\bk} \psi_{\bkp} \, \re^{\ri (\bk + \bkp) \cdot \bT_{ij}}
\nonumber
\\
{} & \quad \times \! \int_0^{\delta t} \!\!\!\! \rd t \!\! \int_0^t \!\! \rd \tp \!\! \int_0^{\tp} \!\!\!\! \rd s \, \re^{\ri \omega_{\bk} t} \, \re^{\ri \omega_{\bkp} s} .
\end{align}
Taking the spatial average ${ \langle \cdot \rangle_{\bT} }$
imposes ${ \bkp \!=\! -\bk }$,
and subsequently ${ \omega_{\bkp} \!=\! -\omega_{\bk} }$ and ${ \psi_{-\bk} \!=\! \psi_{\bk}^* }$. Then, we write
\begin{equation}
\langle A_{i}^{j} \rangle_{\bT} \!=\! \ri m^2 \sum_{\bk} \bk k^2 |\psi_{\bk}|^2 \!\! \int_0^{\delta t} \!\!\!\! \rd t \!\! \int_0^t \!\!\! \rd \tp \!\! \int_0^{\tp} \!\!\! \rd s \, \re^{\ri \omega_{\bk} (t - s)} .
\label{eq:Aij_average}
\end{equation}
The second term $A_{j}^{i}$ contributes an equal amount,
so that
\begin{equation}
\langle A_{i}^{j} \rangle_{\bT} = \langle A_{j}^{i} \rangle_{\bT} .
\label{eq:Aij_vs_Aji}
\end{equation}
We are interested in the dynamics on timescales much longer than a typical dynamical time $\Tdyn$. Hence, we consider the limit ${ \delta t\!\to\! + \infty }$,
with ${ n \!\gg\! 1 }$ fixed.
We now use the identity~\citep[see, e.g.\@,][]{Hamilton2021}
\begin{equation}
\int_0^{\delta t} \!\!\!\! \rd t \!\!\int_0^t \!\! \rd \tp \!\!\int_0^{\tp}\!\!\!\! \rd s \, \re^{\ri x (t - s)} \underset{\delta t \to \infty}{\sim} - \ri \pi \deltaD^{\prime}(x) \, \delta t ,
\label{eq:lim_int_dirac}
\end{equation}
where $\deltaD^{\prime}$ is the derivative of the Dirac delta function.
Substituting this into our expression for ${ \langle A_{i}^{j} \rangle_{\bT} }$ in Eq.~\eqref{eq:Aij_average},
we obtain
\begin{equation}
 \langle A_{i}^{j} \rangle_{\bT} \underset{\delta t \to \infty}{\sim} \pi m^2 \sum_{\bk} |\psi_{\bk}|^2 \bk \, k^{2} \, \deltaD'(\omega_{\bk}) \, \delta t .
\end{equation}
Here, we recognize that the derivative of the delta function can be rewritten as a velocity gradient. Since ${ \p_{\bv_{i}} \deltaD (\omega_{\bk}) \!=\! \bk \, \deltaD' (\omega_{\bk}) }$,
we can rewrite the tensor term as
\begin{equation}
\bk \, k^{2} \deltaD'(\omega_{\bk}) = \frac{\p}{\p \bv_i} \!\cdot\! \big[ (\bk \!\otimes\! \bk) \, \deltaD(\omega_{\bk}) \big] \, .
\end{equation}
This allows us to naturally extract the velocity derivative, yielding the final compact form
\begin{equation}
\langle A_{i}^{j} \rangle_{\bT} \! \underset{\delta t \to \infty}{\sim} \! \pi m^2 \delta t \frac{\p}{\p \bv_i} \!\cdot\! \!\bigg[\! \sum_{\bk} \!|\psi_{\bk}|^2 \, \! \bk \! \otimes \! \bk \, \deltaD(\omega_{\bk}) \!\bigg] \, .
\label{eq:Aij_average_final}
\end{equation}
Combining the results from Eqs.~\eqref{eq:deltaV_A_ij},~\eqref{eq:Aij_vs_Aji} and~\eqref{eq:Aij_average_final}, we finally obtain
\begin{align}
\langle \delta \bv_i^{j} \rangle_{\bT} {} & \underset{\delta t \to \infty}{\sim} 2\pi m^2 \frac{\p}{\p \bv_i} \!\cdot\! \sum_{\bk} |\psi_{\bk}|^2 \bk \!\otimes\! \bk \deltaD[\bk \!\cdot\! \bu_{ij}] \, \delta t
\nonumber
\\
& {} \underset{\delta t \to \infty}{\sim} m^2 \p_{\bv_i} \!\cdot\! \sB(\bu_{ij}) \, \delta t ,
\end{align}
where we recall that ${ \bu_{ij} \!=\! \bv_{i} \!-\! \bv_{j} }$.
Glancing back at Eq.~\eqref{eq:def_b_calc_Rostoker},
we finally obtain the total drift term as
\begin{align}
\bb(\bv_i, \bv_j) &{} = 2\pi m^2 \frac{\p}{\p \bv_i} \!\cdot\! \bigg[ \sum_{\bk} |\psi_{\bk}|^2 \, \bk \!\otimes\! \bk \, \deltaD[\bk \!\cdot\! \bu_{ij}] \bigg] 
\nonumber
\\
& {} = m^2 \frac{\p}{\p \bv_i} \!\cdot\! \sB(\bu_{ij}) .
\end{align}

\subsection{Diffusion term}
\label{app:Diffusion_computation}

We now calculate the diffusion term.
This calculation is simpler than the one for the drift,
since it does not require us to expand
to second order in the perturbation.
From Eq.~\eqref{eq:def_D_Rostoker},
the diffusion term reads
\begin{align}
\sD(\bv_i, \bv_j) & {} = \lim_{\Delta t \to + \infty} \langle \Delta \bv_i^{j} \!\otimes\! \Delta \bv_i^{j}\rangle_{\bT} / \Delta t 
\nonumber
\\
& {} = \lim_{\Delta t \to + \infty} n \, \langle \delta \bv_{i}^{j} \!\otimes\! \delta \bv_i^{j}\rangle_{\bT} / \Delta t.
\end{align}
From Eq.~\eqref{eq:deltav_Rostoker}, averaging over space,
we have
\begin{align}
&\langle \delta \bv_i^{j} \!\otimes\! \delta \bv_i^{j}\rangle_{\bT} = m^2\!\!\int\!\! \frac{\rd \bT_i}{(2\pi)^d} \frac{\rd \bT_j}{(2\pi)^d} \sum_{\bk , \bkp} \bk \!\otimes\! \bkp \psi_{\bk} \psi_{\bkp}
\nonumber
\\
& {} \times \re^{\ri \bT_{ij} \cdot (\bk + \bkp)} \frac{\re^{\ri \bk \cdot \bu_{ij} \delta t} \!-\! 1}{\bk \!\cdot\! \bu_{ij}} \frac{\re^{\ri \bkp \cdot \bu_{ij} \delta t} \!-\! 1}{\bkp \!\cdot\! \bu_{ij}} 
\nonumber
\\
& {} = m^2 \sum_{\bk} \bk \!\otimes\! \bk \, |\psi_{\bk}|^2 \, \bigg|\frac{\re^{\ri \bk \cdot \bu_{ij} \delta t} \!-\! 1}{\bk \!\cdot\! \bu_{ij}} \bigg|^2
\label{eq:calc_diffusion_rostoker}
\\
& {} = m^2 \sum_{\bk} \bk \!\otimes\! \bk \, |\psi_{\bk}|^2 \, \big| \sinc \big[ \half \bk \!\cdot\! \bu_{ij} \delta t \big] \big|^2 \delta t^2 ,
\nonumber
\end{align}
with ${ \sinc(x) \!:=\! \sin(x)/x }$.
By taking the limit of infinitesimal scatter and using the following identity
\begin{equation}
\lim_{\delta t \to + \infty} \delta t \, \big| \sinc \big[ \half x \delta t \big] \big|^2 = 2 \pi \deltaD(x) , 
\end{equation}
we find
\begin{align}
\langle \delta & {} \bv_i^{j} \!\otimes\! \delta \bv_i^{j}\rangle_{\bT} / \delta t = m^2\sum_{\bk} \bk \!\otimes\! \bk \, |\psi_{\bk}|^2
\nonumber
\\
& {} \times \big| \sinc \big[ \half \, \bk \!\cdot\! \bu_{ij} \delta t \big] \big|^2 \delta t
\nonumber
\\
& {} \underset{\delta t \to + \infty}{\longrightarrow} 2\pi m^2\sum_{\bk} \bk \!\otimes\! \bk \, |\psi_{\bk}|^2 \, \deltaD \big[ \bk \!\cdot\! \bu_{ij} \big] .
\end{align}
Using the definition of $\sB$ (Eq.~\ref{eq:def_B_sumK}),
this ultimately leads to
\begin{equation}
\sD(\bv_i, \bv_j) = m^2 \, \sB(\bu_{ij}) .
\end{equation}

\section{Collective effects}
\label{app:Linear_response}

\begin{figure}[htbp!]
\centering
\includegraphics[width=0.45 \textwidth]{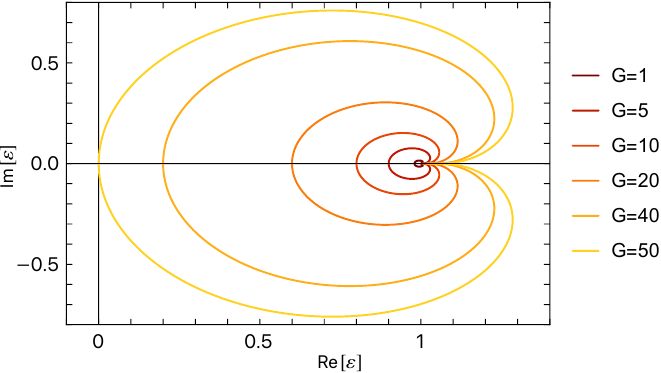}
\caption{Nyquist contour for various values of $G$, evaluated at ${ t \!=\! 0 }$ and with ${ \bk \!=\! (0, \kmin) }$.
We refer to Eq.~\eqref{eq:def_dielectric} for the definition of the dielectric function. For ${ G \!=\! 10 }$, collective effects cannot be neglected.
}
\label{fig:Nyquist}
\end{figure}

In the main text, we focused on the deviations for a system with no collective effects.
In this Appendix,
we briefly explore the case with collective effects,
i.e.\ a situation where the system can support large-scale self-consistent amplifications.
We want to illustrate that \LangevinEP\@ fails to capture deviations in this case.
Generically, the strength of collective effects can be measured via the dielectric function ${ \veps (\bk, \omegaR ) }$
with ${ \omegaR \!\in\! \bbR }$ and
\begin{equation}
\veps (\bk, \omegaR ) = 1 \!-\! \psi_{\bk} \!\! \int \!\! \rd \bv \, \frac{\bk \!\cdot\! \p_{\bv}F(\bv)}{\omegaR \!-\! \bk \!\cdot\! \bv} ,
\label{eq:def_dielectric}
\end{equation}
where the integral is to be interpreted following Landau's prescription~\citep[see, e.g.\@,][]{Nicholson1992}.
The closer ${ \veps ( \bk, \omegaR ) }$ is to unity, the weaker are collective effects. Conversely, if ${ \veps (\bk, \omegaR ) }$ approaches zero,
then the system is inclined to support some self-consistent amplification.
In Fig.~\ref{fig:Nyquist}, we illustrate the Nyquist contour, i.e.\ the curve ${ \omegaR \!\to\! \veps ( \bk, \omegaR ) }$ in the complex plane, for different values of
the gravitational constant, $G$, and ${ \omegaR \!\in\! \bbR }$.

As shown in this figure,
the system for ${ G \!=\! 1 }$ is dynamically hot enough
for collective effects to be negligible.
Therefore, the simulations presented in the main text
remain within the Landau regime.
On the other hand, for ${ G \!=\! 10 }$,
collective effects cannot be neglected anymore,
since ${ \ImPart [\veps] \!\sim\! \RePart[\veps] }$.

\begin{figure}[htbp!]
\centering
\includegraphics[width=0.45 \textwidth]{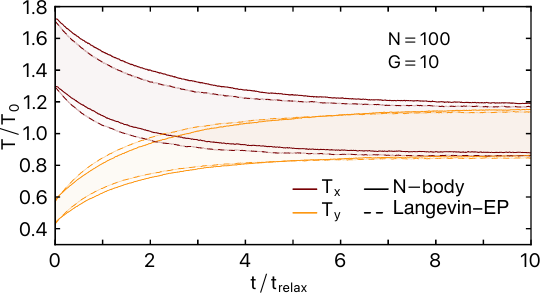}
\caption{Same as Fig.~\ref{fig:Money_plot} but for ${ G \!=\! 10 }$. In the presence of collective effects, \LangevinEP\@ does not match the $N$-body simulations.
}
\label{fig:Contours_G_10}
\end{figure}

In Fig.~\ref{fig:Contours_G_10}, we illustrate the 16\%--84\% contours for ${ G \!=\! 10 }$,
using both $N$-body and \LangevinEP\@ simulations.
As expected, the contours of \LangevinEP\@ do not agree with those of the $N$-body simulations.
This is a numerical illustration that \LangevinEP\@
fails to describe accurately statistical deviations in the presence of collective effects. This is not a surprise.
Indeed, \cite{Feliachi+2022} shows that the \LD\@ Hamiltonian of the Balescu--Lenard equation is not the same as
the one of the Landau equation.
As such, another Langevin equation, 
likely involving non-Gaussian noises,
is needed to capture dynamical deviations
in the presence of collective effects.

\section{Additional validations}
\label{app:Additional_val}

\subsection{Scaling with \texorpdfstring{$N$}{N}} 
\label{app:N_scaling}

\begin{figure}[htbp!]
\centering
\includegraphics[width=0.45 \textwidth]{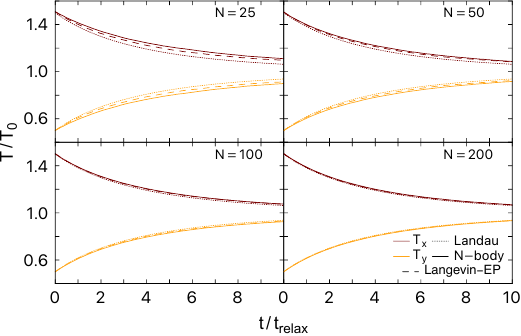}
\includegraphics[width=0.45 \textwidth]{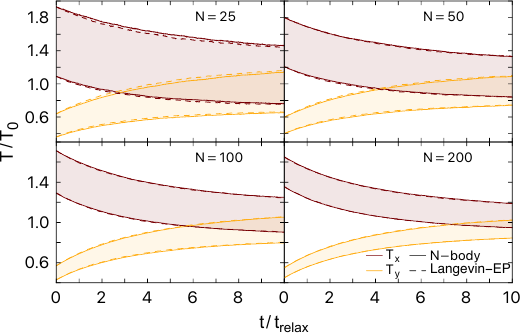}
\includegraphics[width=0.45 \textwidth]{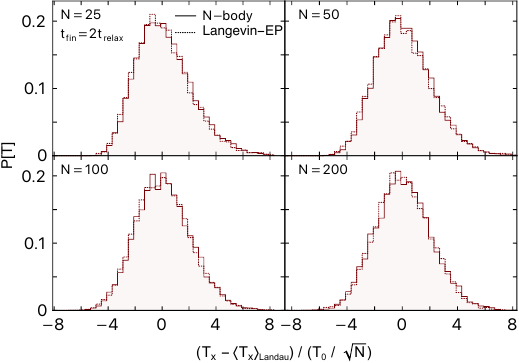}
\caption{Same as Fig.~\ref{fig:Money_plot} (top and middle)
and Fig.~\ref{fig:histograms} (bottom),
for different values of $N$.
\LangevinEP\@ (Eq.~\ref{eq:EP_Langevin})
still matches with the $N$-body simulations
when $N$ is varied.
}
\label{fig:Grid_Histo_N}
\end{figure}

In this Appendix, we revisit Figs.~\ref{fig:Money_plot} and~\ref{fig:histograms}
by varying the total number of particles, $N$.
This is illustrated in Fig.~\ref{fig:Grid_Histo_N}.
As expected, the larger $N$, the tighter the contours,
since the typical fluctuations
around the average scale like $N^{-1/2}$.
For small $N$, the distribution of $T_{x}$
becomes more asymmetric,
but \LangevinEP\@ still matches with the $N$-body approach.
This illustrates how Eq.~\eqref{eq:EP_Langevin}
accurately captures deviations in out-of-equilibrium states.
Similarly, the larger $N$,
the more Gaussian the distribution of temperatures.

\begin{figure}[htbp!]
\centering
\includegraphics[width=0.45 \textwidth]{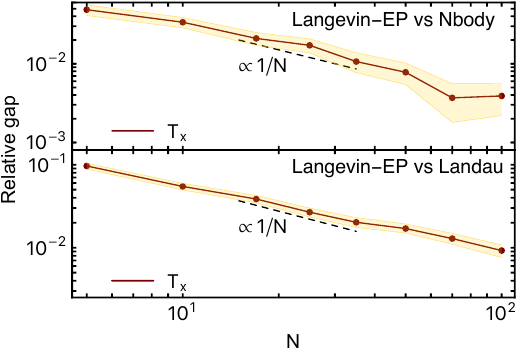}
\caption{Distance ${ \Delta (N) }$ (Eq.~\ref{eq:def_DeltaN}) between the Langevin and $N$-body approaches (top) and Landau (bottom),
as a function of $N$,
using ${ ( \tmin , \tmax ) \!=\! (2 \, \Trelax , 6 \, \Trelax) }$.
Uncertainties are estimated via a bootstrap resampling
over the 10\,000 realizations.
The distance between the three approaches scales like ${1/N}$.
}
\label{fig:Gap_Landau_vs_N}
\end{figure}

Upon further inspection, in the top panel of Fig.~\ref{fig:Grid_Histo_N},
we see that the small discrepancy between the Landau, \LangevinEP\@ and $N$-body computations
is larger for smaller $N$.
To interpret this difference,
we compute the average distance, ${ \Delta (N) }$, via
\begin{equation}
\Delta (N) := \!\! \int_{\tmin}^{\tmax} \!\!\!\! \rd t \, \big| T_{x}^{\LangevinEP}(t) - T_{x}^{\mathrm{Landau}}(t) \big|,
\label{eq:def_DeltaN}
\end{equation}
with ${ \tmin, \tmax \!\propto\! \Trelax }$.
Equation~\eqref{eq:def_DeltaN} can also be
used to estimate the difference between ``\LangevinEP\@'' and ``$N$-body''.
In Fig.~\ref{fig:Gap_Landau_vs_N},
we plot the dependence of ${ \Delta (N) }$,
for $N$ ranging from 5 to 100.
As illustrated in this figure, the discrepancy among the three approaches scales as $1/N$. This scaling originates from two distinct physical approximations. First, both the Landau equation and the \LangevinEP\@ scheme neglect three-body interactions, which naturally occur in the $N$-body dynamics. Second, the discrepancy between the deterministic Landau equation and \LangevinEP\@ arises because the former strictly neglects the fluctuation correlation term $\langle \delta F \delta F \rangle$. This term provides an $\mathcal{O}(1/N)$ correction to the mean-field dynamics, which is intrinsically captured by the stochastic \LangevinEP\@ approach.

\subsection{Fluctuations for cubic observables}
\label{app:Cubic_obs}

\begin{figure}[htbp!]
\centering
\includegraphics[width=0.5 \textwidth]{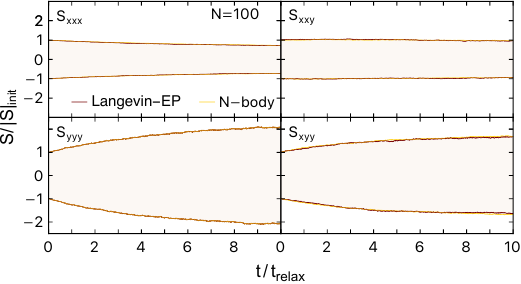}
\caption{Same as Fig.~\ref{fig:Money_plot} but for cubic observables (Eq.~\ref{eq:def_cubic_obs}).
For all the ${S_{\alpha\beta\gamma}}$, the contours of \LangevinEP\@ match with those from $N$-body simulations.
For clarity, we normalised each contour with its initial absolute value.
}
\label{fig:Cubic}
\end{figure}

Instead of looking at quadratic observables like the temperatures $T_{x}$ and $T_{y}$ (see Eq.~\ref{eq:def_Tx})
we can examine cubic observables to determine if the fluctuations remain consistent.
We consider
\begin{equation}
S_{\alpha\beta\gamma} = \sum_{i} v_{i}^{\alpha} v_{i}^{\beta} v_{i}^{\gamma} ,
\label{eq:def_cubic_obs}
\end{equation}
with ${ \alpha,\beta,\gamma }$ associated
with some of the directions, $x$ or $y$.
In Fig.~\ref{fig:Cubic}, we compare the 16\%--84\% contours of the distribution obtained via \LangevinEP\@ and $N$-body simulations.
In this figure, we obtain a very satisfactory agreement
between the \LangevinEP\@ and $N$-body approaches.
This further reinforces our confidence in the fact that
\LangevinEP\@ correctly predicts all the statistics of the deviations.
Compared to Fig.~\ref{fig:Money_plot},
the measurement is noisier in Fig.~\ref{fig:Cubic}.
This was to be expected since we measure
a higher-order moment in $\bv$.
Mitigating this effect would involve increasing
the number of realizations.

\subsection{Door as an initial condition}
\label{app:Door}

In this Appendix, we consider the same numerical setup
as in the main text,
but this time choosing an initial condition given by
\begin{equation}
F(\bv, t \!=\! 0) \!=\! 
\begin{cases} 
\frac{\Mtot}{12\sqrt{T_x T_y}} & \text{if } |v_x|^{2} \!\leq\! 3 T_x \text{ and } |v_y|^{2} \!\leq\! 3 T_y, \\ 
0 & \text{otherwise}.
\end{cases}
\label{eq:DF_door}
\end{equation}
The size of the door is such that initially ${ \langle v_x^2 \rangle \!=\! 3 \, \langle v_{y}^{2} \rangle }$.
\begin{figure}[htbp!]
\centering
\includegraphics[width=0.45 \textwidth]{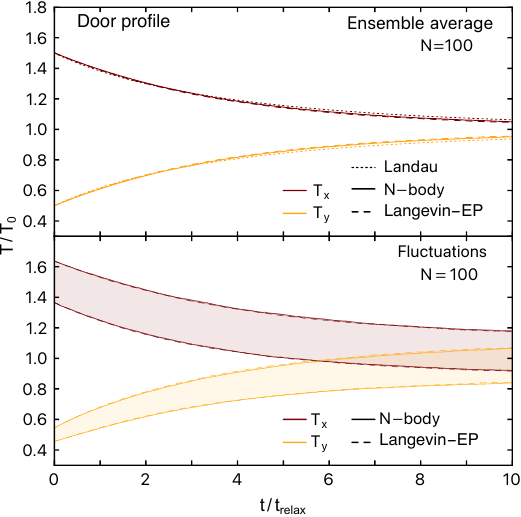}
\caption{Same as Fig.~\ref{fig:Money_plot}
but for the \DF\ from Eq.~\eqref{eq:DF_door}.
The contours and the average behavior are still recovered by \LangevinEP\@. 
}
\label{fig:Moneyplot_DOOR}
\end{figure}

\begin{figure}[htbp!]
\centering
\includegraphics[width=0.5 \textwidth]{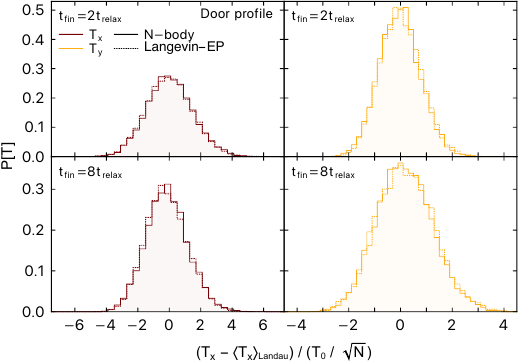}
\caption{Same as Fig.~\ref{fig:histograms}
but for the \DF\ from Eq.~\eqref{eq:DF_door}.
The histograms still agree with the $N$-body.
}
\label{fig:histograms_vs_N_DOOR} 
\end{figure}

As shown in Fig.~\ref{fig:Moneyplot_DOOR},
\LangevinEP\@ describes correctly the deviations for an initial condition that differs significantly from a Maxwellian.
In Fig.~\ref{fig:histograms_vs_N_DOOR}, we consider the histogram
distribution of the temperatures at different times.
We notice that these histograms remain close to Gaussian.
Indeed, their shapes are not related to the shape of the initial \DF\@,
since these histograms show the thermal fluctuations of the system near the average behavior. As $N$ goes to ${ + \infty }$, the shapes of the histograms become more and more Gaussian
because of the Central Limit Theorem.
Similarly, the shape of ${ F(\bv) }$ becomes Gaussian very quickly as the system relaxes to equilibrium.

\begin{figure}[htbp!]
\centering
\includegraphics[width=0.5 \textwidth]{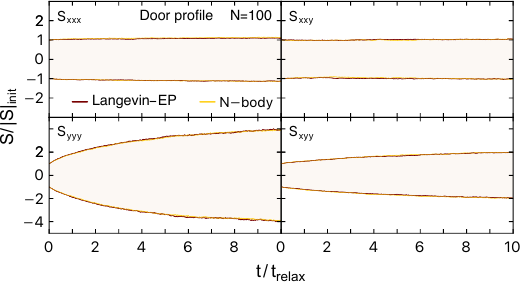}
\caption{Same as Fig.~\ref{fig:Cubic}
but for the \DF\ from Eq.~\eqref{eq:DF_door}.
For all the ${S_{\alpha\beta\gamma}}$, the contours of \LangevinEP\@ match with those from $N$-body simulations.
}
\label{fig:Cubic_DOOR}
\end{figure}

Finally, in Fig.~\ref{fig:Cubic_DOOR}, we consider the fluctuations for cubic observables (Eq.~\ref{eq:def_cubic_obs}).
We recover a very satisfactory match between both approaches.
This is an illustration that our results in the main text
are independent of the initial \DF\@,
provided that it does not sustain strong collective effects,
i.e.\ provided that ${ \veps[F(t \!=\! 0)] \!\simeq\! 1 }$ (see Appendix~\ref{app:Linear_response}).


\begin{thebibliography}{44}
\expandafter\ifx\csname natexlab\endcsname\relax\def\natexlab#1{#1}\fi
\expandafter\ifx\csname bibnamefont\endcsname\relax
  \def\bibnamefont#1{#1}\fi
\expandafter\ifx\csname bibfnamefont\endcsname\relax
  \def\bibfnamefont#1{#1}\fi
\expandafter\ifx\csname citenamefont\endcsname\relax
  \def\citenamefont#1{#1}\fi
\expandafter\ifx\csname url\endcsname\relax
  \def\url#1{\texttt{#1}}\fi
\expandafter\ifx\csname urlprefix\endcsname\relax\def\urlprefix{URL }\fi
\providecommand{\bibinfo}[2]{#2}
\providecommand{\eprint}[2][]{\url{#2}}

\bibitem[{\citenamefont{{Campa} et~al.}(2014)\citenamefont{{Campa}, {Dauxois},
  {Fanelli}, and {Ruffo}}}]{Ruffo+2014}
\bibinfo{author}{\bibfnamefont{A.}~\bibnamefont{{Campa}}},
  \bibinfo{author}{\bibfnamefont{T.}~\bibnamefont{{Dauxois}}},
  \bibinfo{author}{\bibfnamefont{D.}~\bibnamefont{{Fanelli}}},
  \bibnamefont{and} \bibinfo{author}{\bibfnamefont{S.}~\bibnamefont{{Ruffo}}},
  \emph{\bibinfo{title}{{Physics of long-range interacting systems}}}
  (\bibinfo{publisher}{Oxford Univ. Press}, \bibinfo{year}{2014}).

\bibitem[{\citenamefont{{Nicholson}}(1992)}]{Nicholson1992}
\bibinfo{author}{\bibfnamefont{D.~R.} \bibnamefont{{Nicholson}}},
  \emph{\bibinfo{title}{{Introduction to Plasma Theory}}}
  (\bibinfo{publisher}{Krieger}, \bibinfo{year}{1992}).

\bibitem[{\citenamefont{{Binney} and {Tremaine}}(2008)}]{BinneyTremaine2008}
\bibinfo{author}{\bibfnamefont{J.}~\bibnamefont{{Binney}}} \bibnamefont{and}
  \bibinfo{author}{\bibfnamefont{S.}~\bibnamefont{{Tremaine}}},
  \emph{\bibinfo{title}{{Galactic Dynamics: Second Edition}}}
  (\bibinfo{publisher}{Princeton Univ. Press}, \bibinfo{year}{2008}).

\bibitem[{\citenamefont{Heggie and Hut}(2003)}]{Heggie2003}
\bibinfo{author}{\bibfnamefont{D.~C.} \bibnamefont{Heggie}} \bibnamefont{and}
  \bibinfo{author}{\bibfnamefont{P.}~\bibnamefont{Hut}},
  \emph{\bibinfo{title}{The gravitational million-body problem}}
  (\bibinfo{publisher}{Cambridge Univ. Press}, \bibinfo{year}{2003}).

\bibitem[{\citenamefont{{Bouchet} and {Venaille}}(2012)}]{Bouchet+2012}
\bibinfo{author}{\bibfnamefont{F.}~\bibnamefont{{Bouchet}}} \bibnamefont{and}
  \bibinfo{author}{\bibfnamefont{A.}~\bibnamefont{{Venaille}}},
  \bibinfo{journal}{Phys. Rep.} \textbf{\bibinfo{volume}{515}},
  \bibinfo{pages}{227} (\bibinfo{year}{2012}).

\bibitem[{\citenamefont{{Dauxois} et~al.}(2002)\citenamefont{{Dauxois},
  {Ruffo}, {Arimondo}, and {Wilkens}}}]{Dauxois+2002}
\bibinfo{author}{\bibfnamefont{T.}~\bibnamefont{{Dauxois}}},
  \bibinfo{author}{\bibfnamefont{S.}~\bibnamefont{{Ruffo}}},
  \bibinfo{author}{\bibfnamefont{E.}~\bibnamefont{{Arimondo}}},
  \bibnamefont{and}
  \bibinfo{author}{\bibfnamefont{M.}~\bibnamefont{{Wilkens}}}, in
  \emph{\bibinfo{booktitle}{Dynamics and Thermodynamics of Systems with
  Long-Range Interactions}} (\bibinfo{publisher}{Springer},
  \bibinfo{year}{2002}), vol. \bibinfo{volume}{602}, pp.
  \bibinfo{pages}{1--19}.

\bibitem[{\citenamefont{{Klimontovich}}(1967)}]{Klimontovich1967}
\bibinfo{author}{\bibfnamefont{Y.~L.} \bibnamefont{{Klimontovich}}},
  \emph{\bibinfo{title}{{The Statistical Theory of Non-Equilibrium Processes in
  a Plasma}}} (\bibinfo{publisher}{Elsevier}, \bibinfo{year}{1967}).

\bibitem[{\citenamefont{Balescu}(1997)}]{Balescu1997}
\bibinfo{author}{\bibfnamefont{R.}~\bibnamefont{Balescu}},
  \emph{\bibinfo{title}{{Statistical Dynamics: Matter out of Equilibrium}}}
  (\bibinfo{publisher}{Imperial Coll., London}, \bibinfo{year}{1997}).

\bibitem[{\citenamefont{{Lifshitz} and {Pitaevskii}}(1981)}]{Landau+1981}
\bibinfo{author}{\bibfnamefont{E.~M.} \bibnamefont{{Lifshitz}}}
  \bibnamefont{and} \bibinfo{author}{\bibfnamefont{L.~P.}
  \bibnamefont{{Pitaevskii}}}, \emph{\bibinfo{title}{{Physical kinetics}}}
  (\bibinfo{publisher}{Pergamon Press}, \bibinfo{year}{1981}).

\bibitem[{\citenamefont{{Hamilton} and {Fouvry}}(2024)}]{Fouvry+2024}
\bibinfo{author}{\bibfnamefont{C.}~\bibnamefont{{Hamilton}}} \bibnamefont{and}
  \bibinfo{author}{\bibfnamefont{J.-B.} \bibnamefont{{Fouvry}}},
  \bibinfo{journal}{Phys. Plasmas} \textbf{\bibinfo{volume}{31}},
  \bibinfo{eid}{120901} (\bibinfo{year}{2024}).

\bibitem[{\citenamefont{{Landau}}(1936)}]{Landau1936}
\bibinfo{author}{\bibfnamefont{L.~D.} \bibnamefont{{Landau}}},
  \bibinfo{journal}{Phys. Z. Sowj. Union} \textbf{\bibinfo{volume}{10}},
  \bibinfo{pages}{154} (\bibinfo{year}{1936}).

\bibitem[{\citenamefont{{Lynden-Bell} and {Wood}}(1968)}]{Lynden+1968}
\bibinfo{author}{\bibfnamefont{D.}~\bibnamefont{{Lynden-Bell}}}
  \bibnamefont{and} \bibinfo{author}{\bibfnamefont{R.}~\bibnamefont{{Wood}}},
  \bibinfo{journal}{MNRAS} \textbf{\bibinfo{volume}{138}}, \bibinfo{pages}{495}
  (\bibinfo{year}{1968}).

\bibitem[{\citenamefont{{Roule} et~al.}(2025)\citenamefont{{Roule}, {Fouvry},
  {Pichon}, and {Chavanis}}}]{Roule+2025}
\bibinfo{author}{\bibfnamefont{M.}~\bibnamefont{{Roule}}},
  \bibinfo{author}{\bibfnamefont{J.-B.} \bibnamefont{{Fouvry}}},
  \bibinfo{author}{\bibfnamefont{C.}~\bibnamefont{{Pichon}}}, \bibnamefont{and}
  \bibinfo{author}{\bibfnamefont{P.-H.} \bibnamefont{{Chavanis}}},
  \bibinfo{journal}{A\&A} \textbf{\bibinfo{volume}{699}}, \bibinfo{eid}{A140}
  (\bibinfo{year}{2025}).

\bibitem[{\citenamefont{{Asano} and {Portegies Zwart}}(2026)}]{Asano+2026}
\bibinfo{author}{\bibfnamefont{T.}~\bibnamefont{{Asano}}} \bibnamefont{and}
  \bibinfo{author}{\bibfnamefont{S.}~\bibnamefont{{Portegies Zwart}}},
  \bibinfo{journal}{arXiv} \bibinfo{eid}{2604.12053} (\bibinfo{year}{2026}).

\bibitem[{\citenamefont{{Touchette}}(2009)}]{Touchette2009}
\bibinfo{author}{\bibfnamefont{H.}~\bibnamefont{{Touchette}}},
  \bibinfo{journal}{Phys. Rep.} \textbf{\bibinfo{volume}{478}},
  \bibinfo{pages}{1} (\bibinfo{year}{2009}).

\bibitem[{\citenamefont{{Feliachi} and {Bouchet}}(2021)}]{Feliachi+2021}
\bibinfo{author}{\bibfnamefont{O.}~\bibnamefont{{Feliachi}}} \bibnamefont{and}
  \bibinfo{author}{\bibfnamefont{F.}~\bibnamefont{{Bouchet}}},
  \bibinfo{journal}{J. Stat. Phys.} \textbf{\bibinfo{volume}{183}},
  \bibinfo{eid}{42} (\bibinfo{year}{2021}).

\bibitem[{\citenamefont{{Feliachi} and {Bouchet}}(2022)}]{Feliachi+2022}
\bibinfo{author}{\bibfnamefont{O.}~\bibnamefont{{Feliachi}}} \bibnamefont{and}
  \bibinfo{author}{\bibfnamefont{F.}~\bibnamefont{{Bouchet}}},
  \bibinfo{journal}{J. Stat. Phys.} \textbf{\bibinfo{volume}{186}},
  \bibinfo{eid}{22} (\bibinfo{year}{2022}).

\bibitem[{\citenamefont{{Feliachi} and {Fouvry}}(2024)}]{Feliachi+2024}
\bibinfo{author}{\bibfnamefont{O.}~\bibnamefont{{Feliachi}}} \bibnamefont{and}
  \bibinfo{author}{\bibfnamefont{J.-B.} \bibnamefont{{Fouvry}}},
  \bibinfo{journal}{Phys. Rev. E} \textbf{\bibinfo{volume}{110}},
  \bibinfo{eid}{024108} (\bibinfo{year}{2024}).

\bibitem[{\citenamefont{{Fontbona} et~al.}(2009)\citenamefont{{Fontbona},
  {Gu\'erin}, and {M\'el\'eard}}}]{Fontbona+2009}
\bibinfo{author}{\bibfnamefont{J.}~\bibnamefont{{Fontbona}}},
  \bibinfo{author}{\bibfnamefont{H.}~\bibnamefont{{Gu\'erin}}},
  \bibnamefont{and}
  \bibinfo{author}{\bibfnamefont{S.}~\bibnamefont{{M\'el\'eard}}},
  \bibinfo{journal}{Probab. Theory Relat. Fields}
  \textbf{\bibinfo{volume}{143}}, \bibinfo{pages}{329} (\bibinfo{year}{2009}).

\bibitem[{\citenamefont{{Fu} et~al.}(2025)\citenamefont{{Fu}, {Angus}, {Qin},
  and {Geyko}}}]{Fu+2025}
\bibinfo{author}{\bibfnamefont{Y.}~\bibnamefont{{Fu}}},
  \bibinfo{author}{\bibfnamefont{J.~R.} \bibnamefont{{Angus}}},
  \bibinfo{author}{\bibfnamefont{H.}~\bibnamefont{{Qin}}}, \bibnamefont{and}
  \bibinfo{author}{\bibfnamefont{V.~I.} \bibnamefont{{Geyko}}},
  \bibinfo{journal}{Phys. Rev. E} \textbf{\bibinfo{volume}{111}},
  \bibinfo{eid}{025211} (\bibinfo{year}{2025}).

\bibitem[{Sup()}]{SupplMat}
\bibinfo{note}{See Supplemental Material.}

\bibitem[{\citenamefont{{Chavanis}}(2013)}]{Chavanis2013}
\bibinfo{author}{\bibfnamefont{P.-H.} \bibnamefont{{Chavanis}}},
  \bibinfo{journal}{A\&A} \textbf{\bibinfo{volume}{556}}, \bibinfo{eid}{A93}
  (\bibinfo{year}{2013}).

\bibitem[{\citenamefont{{Bouchet} et~al.}(2016)\citenamefont{{Bouchet},
  {Grafke}, {Tangarife}, and {Vanden-Eijnden}}}]{BouchetVanden2016}
\bibinfo{author}{\bibfnamefont{F.}~\bibnamefont{{Bouchet}}},
  \bibinfo{author}{\bibfnamefont{T.}~\bibnamefont{{Grafke}}},
  \bibinfo{author}{\bibfnamefont{T.}~\bibnamefont{{Tangarife}}},
  \bibnamefont{and}
  \bibinfo{author}{\bibfnamefont{E.}~\bibnamefont{{Vanden-Eijnden}}},
  \bibinfo{journal}{J. Stat. Phys.} \textbf{\bibinfo{volume}{162}},
  \bibinfo{pages}{793} (\bibinfo{year}{2016}).

\bibitem[{\citenamefont{{Bertini} et~al.}(2015)\citenamefont{{Bertini}, {De
  Sole}, {Gabrielli}, {Jona-Lasinio}, and {Landim}}}]{Roma2015}
\bibinfo{author}{\bibfnamefont{L.}~\bibnamefont{{Bertini}}},
  \bibinfo{author}{\bibfnamefont{A.}~\bibnamefont{{De Sole}}},
  \bibinfo{author}{\bibfnamefont{D.}~\bibnamefont{{Gabrielli}}},
  \bibinfo{author}{\bibfnamefont{G.}~\bibnamefont{{Jona-Lasinio}}},
  \bibnamefont{and} \bibinfo{author}{\bibfnamefont{C.}~\bibnamefont{{Landim}}},
  \bibinfo{journal}{Rev. Mod. Phys.} \textbf{\bibinfo{volume}{87}},
  \bibinfo{pages}{593} (\bibinfo{year}{2015}).

\bibitem[{\citenamefont{{Risken}}(1989)}]{Risken1989}
\bibinfo{author}{\bibfnamefont{H.}~\bibnamefont{{Risken}}},
  \emph{\bibinfo{title}{{The Fokker-Planck equation}}}
  (\bibinfo{publisher}{Springer}, \bibinfo{year}{1989}).

\bibitem[{\citenamefont{{Heyvaerts} et~al.}(2017)\citenamefont{{Heyvaerts},
  {Fouvry}, {Chavanis}, and {Pichon}}}]{Heyvaerts+2017}
\bibinfo{author}{\bibfnamefont{J.}~\bibnamefont{{Heyvaerts}}},
  \bibinfo{author}{\bibfnamefont{J.-B.} \bibnamefont{{Fouvry}}},
  \bibinfo{author}{\bibfnamefont{P.-H.} \bibnamefont{{Chavanis}}},
  \bibnamefont{and} \bibinfo{author}{\bibfnamefont{C.}~\bibnamefont{{Pichon}}},
  \bibinfo{journal}{MNRAS} \textbf{\bibinfo{volume}{469}},
  \bibinfo{pages}{4193} (\bibinfo{year}{2017}).

\bibitem[{\citenamefont{{Rostoker}}(1964{\natexlab{a}})}]{Rostoker1964a}
\bibinfo{author}{\bibfnamefont{N.}~\bibnamefont{{Rostoker}}},
  \bibinfo{journal}{Phys. Fluids} \textbf{\bibinfo{volume}{7}},
  \bibinfo{pages}{479} (\bibinfo{year}{1964}{\natexlab{a}}).

\bibitem[{\citenamefont{{Rostoker}}(1964{\natexlab{b}})}]{Rostoker1964b}
\bibinfo{author}{\bibfnamefont{N.}~\bibnamefont{{Rostoker}}},
  \bibinfo{journal}{Phys. Fluids} \textbf{\bibinfo{volume}{7}},
  \bibinfo{pages}{491} (\bibinfo{year}{1964}{\natexlab{b}}).

\bibitem[{\citenamefont{{Du} et~al.}(2025)\citenamefont{{Du}, {Li}, {Xie}, and
  {Yu}}}]{Kai+2025}
\bibinfo{author}{\bibfnamefont{K.}~\bibnamefont{{Du}}},
  \bibinfo{author}{\bibfnamefont{L.}~\bibnamefont{{Li}}},
  \bibinfo{author}{\bibfnamefont{Y.}~\bibnamefont{{Xie}}}, \bibnamefont{and}
  \bibinfo{author}{\bibfnamefont{Y.}~\bibnamefont{{Yu}}}, \bibinfo{journal}{J.
  Comp. Phys.} \textbf{\bibinfo{volume}{543}}, \bibinfo{eid}{114387}
  (\bibinfo{year}{2025}).

\bibitem[{\citenamefont{{Bouchet} et~al.}(2013)\citenamefont{{Bouchet},
  {Nardini}, and {Tangarife}}}]{Bouchet+2013}
\bibinfo{author}{\bibfnamefont{F.}~\bibnamefont{{Bouchet}}},
  \bibinfo{author}{\bibfnamefont{C.}~\bibnamefont{{Nardini}}},
  \bibnamefont{and}
  \bibinfo{author}{\bibfnamefont{T.}~\bibnamefont{{Tangarife}}},
  \bibinfo{journal}{J. Stat. Phys.} \textbf{\bibinfo{volume}{153}},
  \bibinfo{pages}{572} (\bibinfo{year}{2013}).

\bibitem[{\citenamefont{{Dean}}(1996)}]{Dean1996}
\bibinfo{author}{\bibfnamefont{D.~S.} \bibnamefont{{Dean}}},
  \bibinfo{journal}{J. Phys. A} \textbf{\bibinfo{volume}{29}},
  \bibinfo{pages}{L613} (\bibinfo{year}{1996}).

\bibitem[{\citenamefont{{Kawasaki}}(1998)}]{Kawazaki1998}
\bibinfo{author}{\bibfnamefont{K.}~\bibnamefont{{Kawasaki}}},
  \bibinfo{journal}{J. Stat. Phys.} \textbf{\bibinfo{volume}{93}},
  \bibinfo{pages}{527} (\bibinfo{year}{1998}).

\bibitem[{\citenamefont{{Cornalba} and {Fischer}}(2023)}]{Cornalba+2023}
\bibinfo{author}{\bibfnamefont{F.}~\bibnamefont{{Cornalba}}} \bibnamefont{and}
  \bibinfo{author}{\bibfnamefont{J.}~\bibnamefont{{Fischer}}},
  \bibinfo{journal}{Arch. Ration. Mech. Anal.} \textbf{\bibinfo{volume}{247}},
  \bibinfo{eid}{76} (\bibinfo{year}{2023}).

\bibitem[{\citenamefont{{Balescu}}(1960)}]{Balescu1960}
\bibinfo{author}{\bibfnamefont{R.}~\bibnamefont{{Balescu}}},
  \bibinfo{journal}{Phys. Fluids} \textbf{\bibinfo{volume}{3}},
  \bibinfo{pages}{52} (\bibinfo{year}{1960}).

\bibitem[{\citenamefont{{Lenard}}(1960)}]{Lenard1960}
\bibinfo{author}{\bibfnamefont{A.}~\bibnamefont{{Lenard}}},
  \bibinfo{journal}{Ann. Phys. (N.-Y.)} \textbf{\bibinfo{volume}{10}},
  \bibinfo{pages}{390} (\bibinfo{year}{1960}).

\bibitem[{git()}]{github}
\bibinfo{howpublished}{\url{https://github.com/anwar-elrhirhayi/Langevin_Landau}}.

\bibitem[{\citenamefont{{Hamilton}}(2021)}]{Hamilton2021}
\bibinfo{author}{\bibfnamefont{C.}~\bibnamefont{{Hamilton}}},
  \bibinfo{journal}{MNRAS} \textbf{\bibinfo{volume}{501}},
  \bibinfo{pages}{3371} (\bibinfo{year}{2021}).

\bibitem[{\citenamefont{Bhattacharya and Waymire}(2021)}]{Bhattacharya2021}
\bibinfo{author}{\bibfnamefont{R.}~\bibnamefont{Bhattacharya}}
  \bibnamefont{and} \bibinfo{author}{\bibfnamefont{E.~C.}
  \bibnamefont{Waymire}}, \emph{\bibinfo{title}{The Functional Central Limit
  Theorem (FCLT)}} (\bibinfo{publisher}{Springer}, \bibinfo{year}{2021}).

\bibitem[{\citenamefont{Hairer et~al.}(2006)\citenamefont{Hairer, Lubich, and
  Wanner}}]{Hairer+2006}
\bibinfo{author}{\bibfnamefont{E.}~\bibnamefont{Hairer}},
  \bibinfo{author}{\bibfnamefont{C.}~\bibnamefont{Lubich}}, \bibnamefont{and}
  \bibinfo{author}{\bibfnamefont{G.}~\bibnamefont{Wanner}},
  \emph{\bibinfo{title}{{Geometric numerical integration: Second Edition}}}
  (\bibinfo{publisher}{Springer}, \bibinfo{year}{2006}).

\bibitem[{\citenamefont{{Yoshida}}(1990)}]{Yoshida1990}
\bibinfo{author}{\bibfnamefont{H.}~\bibnamefont{{Yoshida}}},
  \bibinfo{journal}{Phys. Lett. A} \textbf{\bibinfo{volume}{150}},
  \bibinfo{pages}{262} (\bibinfo{year}{1990}).

\bibitem[{\citenamefont{Kloeden and Platen}(2011)}]{Kloeden+2011}
\bibinfo{author}{\bibfnamefont{P.}~\bibnamefont{Kloeden}} \bibnamefont{and}
  \bibinfo{author}{\bibfnamefont{E.}~\bibnamefont{Platen}},
  \emph{\bibinfo{title}{Numerical Solution of Stochastic Differential
  Equations}} (\bibinfo{publisher}{Springer}, \bibinfo{year}{2011}).

\bibitem[{\citenamefont{Hong and Sun}(2022)}]{Hong+2022}
\bibinfo{author}{\bibfnamefont{J.}~\bibnamefont{Hong}} \bibnamefont{and}
  \bibinfo{author}{\bibfnamefont{L.}~\bibnamefont{Sun}},
  \emph{\bibinfo{title}{Stochastic Structure-Preserving Numerical Methods}}
  (\bibinfo{publisher}{Springer}, \bibinfo{year}{2022}).

\bibitem[{\citenamefont{{Zhang} et~al.}(2020)\citenamefont{{Zhang}, {Fu}, and
  {Qin}}}]{Zhang+2020}
\bibinfo{author}{\bibfnamefont{X.}~\bibnamefont{{Zhang}}},
  \bibinfo{author}{\bibfnamefont{Y.}~\bibnamefont{{Fu}}}, \bibnamefont{and}
  \bibinfo{author}{\bibfnamefont{H.}~\bibnamefont{{Qin}}},
  \bibinfo{journal}{\pre} \textbf{\bibinfo{volume}{102}}
  (\bibinfo{year}{2020}).

\bibitem[{\citenamefont{{Bouchet}}(2020)}]{Bouchet2020}
\bibinfo{author}{\bibfnamefont{F.}~\bibnamefont{{Bouchet}}},
  \bibinfo{journal}{J. Stat. Phys.} \textbf{\bibinfo{volume}{181}},
  \bibinfo{pages}{515} (\bibinfo{year}{2020}).

\end{thebibliography}
\end{document}